\documentclass[nature, superscriptaddress, longbibliography, onecolumn]{revtex4-2}
\usepackage{amsmath}
\usepackage{graphicx}
\usepackage{dcolumn}    
\usepackage{bm}     
\usepackage{epstopdf}
\usepackage{graphicx}
\usepackage{stmaryrd}
\usepackage{tikz}
\usepackage{lipsum}
\usepackage{pgf}
\usepackage{float}
\usepackage{hyperref}
\usepackage{tabularx}
\usepackage{subfigure}
\usepackage{physics}
\usepackage{float}
\usepackage{subfigure}

\definecolor{lightblue}{RGB}{0,170,255}

\usepackage[normalem]{ulem}
\newcommand{\dps}{\displaystyle}

\newcommand{\om}{\iffalse}
\newcommand{\pd}[2]{\frac{\partial #1}{\partial #2}}

\newcommand{\ba}{\arraycolsep 0.3ex \begin{array}{rl}}
	\newcommand{\ea}{\end{array}}
\newcommand{\bc}{\begin{cases}}
	\newcommand{\ec}{\end{cases}}
\usepackage{color}
\newcolumntype{C}[1]{>{\centering\arraybackslash}p{#1}}
\begin{document}
\title{Non-linear ballistic response of quantum spin-Hall edge states}

\author{Pankaj Bhalla*}
\affiliation{Beijing Computational Science Research Centre, Beijing, 100193, China}
\affiliation{ARC Centre of Excellence in Future Low-Energy Electronics Technologies, The University of New South Wales, Sydney 2052, Australia}
\affiliation{School of Physics, The University of New South Wales, Sydney 2052, Australia}

\author{Ming-Xun Deng}
\affiliation{Guangdong Provincial Key Laboratory of Quantum Engineering and Quantum Materials,  Guangzhou 510006, China}
\affiliation{South China Normal University, Guangzhou 510006, China}

\author{Rui-Qiang Wang}
\affiliation{Guangdong Provincial Key Laboratory of Quantum Engineering and Quantum Materials, Guangzhou 510006, China}
\affiliation{South China Normal University, Guangzhou 510006, China}

\author{Lan Wang}
\affiliation{School of Science, RMIT University, Melbourne, VIC 3000, Australia}
\affiliation{ARC Centre of Excellence in Future Low-Energy Electronics Technologies, RMIT Node, RMIT University, Melbourne, VIC 3000, Australia}

\author{Dimitrie Culcer}
\affiliation{School of Physics, The University of New South Wales, Sydney 2052, Australia}
\affiliation{ARC Centre of Excellence in Future Low-Energy Electronics Technologies, The University of New South Wales, Sydney 2052, Australia}	

\date{\today}
\begin{abstract}
\begin{center}
    \textbf{Abstract}
\end{center}
Topological edge states exhibit dissipationless transport and electrically-driven topological phase transitions, making them ideal for next-generation transistors that are not constrained by Moore’s law. Nevertheless, their dispersion has never been probed and is often assumed to be simply linear, without any rigorous justification. Here we determine the non-linear electrical response of topological edge states in the ballistic regime and demonstrate the way this response ascertains the presence of symmetry breaking terms in the edge dispersion, such as deviations from non-linearity and tilted spin quantization axes. The non-linear response stems from discontinuities in the band occupation on either side of a Zeeman gap, and its direction is set by the spin orientation with respect to the Zeeman field. We determine the edge dispersion for several classes of topological materials and discuss experimental measurement.
\end{abstract}
\maketitle

\section*{Introduction} Topological materials such as topological insulators, Weyl semimetals, and transition metal dichalcogenides, are novel quantum materials hosting helical or chiral spin-momentum locked states on their surfaces and edges \cite{zhang_NP2009, hasan_RMP2010, liu_PRB2010, chu_PRB2011, adagideli_PRL2005, yan_RPP2012, Liu_Sc2014, weber_NN2014, farrell_PRB2016, yan_ARCMP2017, wang_NM2017, he_ARCM2018, shen_book}, which may enable dissipationless transport. This fact, coupled with the possibility of electrically-driven topological phase transitions \cite{wang_Nat2016, collins_NP2018, scharf_PRB2015, rostami_PRR2020} has led to an explosion of interest in topological edge state transistors as novel, power-saving building blocks for next-generation integrated circuits \cite{nemnes_JAP2004, fabian_PRB2004, huang_APL2007, wray_Nat2012, vicarelli_NM2012, cheng_SR2016, collins_NP2018}. The first step in this road map is achieving reliable ballistic samples, in which the electron mean free path $l$ is much greater than the length $d$ of the channel. The corresponding conduction picture is frequently described by the Landauer-Buttiker formalism \cite{landauer_ZPB1987, larkin_JETP1986, bagwell_PRB1989, meir_PRL1992, landauer_PhyScr1992, datta_book1993, datta_book, green_JPCM2000, liang_Nat2001, Jeong_JAP2010, knez_PRL2011, vadim_PRL2013, wang_JNS2014, knez_PRL2014, Du_PRL2015, Li_PRL2015, fei_NatPhy2017}.

The ballistic regime can exhibit linear as well as non-linear transport, as observed in quantum point contacts, three-terminal ballistic branches, asymmetric micro-junctions and related structures \cite{khmel_PS1986, christen_EPL1996, song_PRL1998, buttiker_book1998, you_PRB2000, shorubalko_APL2001, reitzenstein_PRL2002, polianski_PRB2007, safi_PRB2011, chang_PRL2015, korniyenko_PRB2016, rostami_PRB2018, texier_PRB2018, mardanya_PRB2018, ildarabadi_PRB2021}. The transport properties of conventional devices such as quantum point contacts are typically tailored by device geometry in a similar fashion to the transmission properties of a waveguide \cite{bennakker_SSP1991}. On the other hand, topological edge states are expected to exhibit inversion symmetry breaking terms intrinsic to the edge Hamiltonian, which itself should enable a non-linear electrical response in the technologically relevant ballistic regime, without additional structure inversion symmetry built into the device. The ballistic regime is interesting for fundamental reasons as well, since in the absence of disorder the edge states have an unambiguous fingerprint. Nevertheless this non-linear phenomenon has not been considered to date, in fact the presence of inversion-symmetry breaking terms in the dispersion has never been probed, because the standard tools for this, angle-resolved photoemission and scanning tunnelling microscopy, do not work for single edges. This knowledge gap motivates us to develop here a quantum kinetic theory for the non-linear response of ballistic topological edge states. We focus on the simplest, but experimentally most relevant, case of a single channel with perfect transmission to the contacts. 

\section*{Results}

\begin{figure}[t!]
     \centering
     \includegraphics[height=4cm,width=8cm]{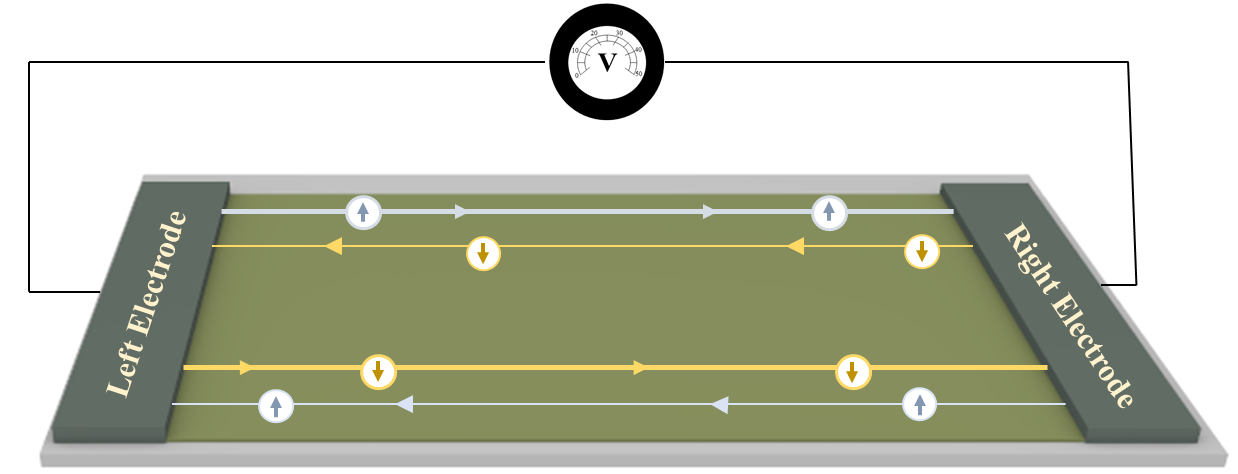}
\caption{\textbf{Experimental setup for the measurement of the non-linear current}. In the setup shown the voltages are measured along the left and right electrodes. Here the edge states occur in pairs having spin up (blue) and down (yellow). Thick lines refer to occupied states and thin lines to unoccupied states.}
\label{fig:expt}
\end{figure}

The central result of our work is the following generic expression for the non-linear contribution to the current 
\begin{equation}
    \ba
   j^{(2)} &\dps = (e^3/h)(V_L - V_R)^2 \big[f'(E_Z-\mu) - f'(E_Z+\mu) \big].
   \label{eqn:NLCexpression}
    \ea
\end{equation}
Here $f'(E_Z-\mu)$ is the derivative of the Fermi Dirac distribution function $[1+e^{\beta(E_Z - \mu)}]^{-1}$ with respect to the Zeeman energy $E_Z$, with $\beta = 1/(k_BT)$, $k_B$ the Boltzmann constant, $T$ the absolute temperature, while $\mu$ refers to the chemical potential, and $V_L$ ($V_R$) is the potential of the left (right) electrode. 
\begin{figure}[!t]
    \centering
    \includegraphics[width=8cm, height=6cm]{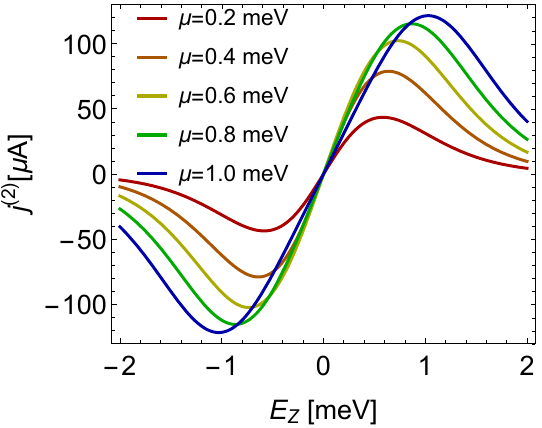}
    \caption{\textbf{Non-linear response of topological edge states}. The plot shows the second-order electrical current due to quantum spin-Hall edge states in the ballistic regime in response to a potential difference between the left and the right electrodes at a finite Zeeman energy $E_Z$ (meV) for different chemical potentials $\mu$ (meV) at a fixed low temperature T = 5 K.}
    \label{fig:NLR_main}
\end{figure}
Moreover, it is nonzero only due to the asymmetry created by the Zeeman energy as illustrated in Fig.~\ref{fig:NLR_main}. Here we have plotted the second-order current as a function of the Zeeman energy and the chemical potential referring the experimental setup shown in Fig.~\ref{fig:expt}. It is observed that $j^{(2)}$ follows the smeared $\delta$-function shape due to the wave vector derivative of the Fermi function. Note that for vanishing Zeeman energy, the current becomes zero. The response is unidirectional, with the direction set by the spin orientation with respect to the magnetic field, and has opposite signs on the two edges. The non-linear part of the current changes sign on reversing the direction of the Zeeman term. The necessity of a Zeeman field reflects the fact that, beyond the linear regime, the role of time-reversal symmetry is non-trivial \cite{altshuler_JETP1985, webb_PRB1988, buttiker_JPCM1993, thakur_IJMPB2004, david_PRL2004, lofgren_PRL2004, datta_book2, shi_PRB2019, sun_PRB2020}. It is consistent with the recent finding that a non-linear reciprocal current requires time-reversal symmetry breaking either by introducing the magnetic order at the microscopic level or by incorporating the irreversibility at the macroscopic \cite{morimoto_SR2018}. In the presence of time-reversal symmetry, the symmetrical shift in the electronic band structure in the valence and conduction bands due to the potential gradient between two electrodes nullifies the net motion of the carriers, giving zero current \cite{morimoto_SR2018, arpit_PRB2020}.

In order to make concrete experimental predictions, below we evaluate the ballistic non-linear edge response for several classes of topological materials such as Na$_3$Bi, Bi$_2$Se$_3$, HgTe, WTe$_2$ and quantum anomalous Hall edge states. Our main conclusions are: (i) Unlike linear response, the non-linear response can ascertain the presence of symmetry-breaking terms in the dispersion, an important step forward considering that the dispersion of topological edge states has never been probed; (ii) the shape of the non-linear response as a function of chemical potential and Zeeman energy does not depend on the details of the band structure. Yet the response only occurs if mirror-symmetry breaking terms are present in the band structure. Hence if a ballistic non-linear response exists, it has the shape of Fig.~\ref{fig:NLR_main}; (iii) Physically, the non-linear response arises when there is a discontinuity in the linear current. To detect a finite non-linear electrical response one straightforwardly tunes the chemical potential through the Zeeman gap, whether at the origin or at finite wave vector, while monitoring the voltage at twice the applied frequency. (iv) Although Zitterbewegung terms are formally present in the response we find they have no physical consequences in one dimension, and the Berry connection does not ultimately play a role in ballistic transport.

\begin{figure}[t!]
     \centering
     \includegraphics[height=6cm,width=7cm]{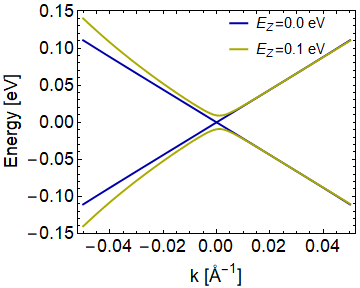}
\caption{\textbf{Edge energy dispersion for Bi$_2$Se$_3$}. We have included the Zeeman energy term $E_Z$ as well as the term cubic in wave vector, which is the descendant of the \textit{warping term} in 2D topological insulators. We have considered a fixed warping coefficient $\lambda = 100$ eV$\AA^3$.}
\label{fig:dispersion}
\end{figure}

We consider a generic edge of finite length $d$ which is described by the total Hamiltonian $H = H_0 + H_Z + V(x)$, where $H_0$ is the Bloch band Hamiltonian corresponding to the edge dispersion of system under consideration, $H_Z$ is the Zeeman energy term, and $V(x)$ the applied potential difference. We briefly sketch the derivation for the edge state Hamiltonian for Bi$_2$Se$_3$. We start with the effective 4 $\times$ 4 Hamiltonian for Bi$_2$Se$_3$ in the basis $\{ \psi_{c \uparrow}, \psi_{v \uparrow}, \psi_{c \downarrow}, \psi_{v \downarrow} \}$ which can be expressed in the block diagonal form as
\begin{equation}
\ba
H &\dps = \left(
\begin{array}{cc}
h_+({\bm k}) & 0 \\
0 & h_-({\bm k})
\end{array}
\right).
\ea
\end{equation}
Here the block matrices $h_{\pm} = E_0\mathcal{I} + D_0(k_x^2 + k_y^2) \mathcal{I} + \lambda (k_x^3 - 3k_x k_y^2) \sigma_z - \alpha k_{\mp} \sigma_y$ where the second term which is quadratic term in the wave vector refers to the kinetic energy term having $D_0$ as material dependent parameter, third term is for the warping which reduces the infinite mirror planes to three \cite{Fu_PRL2009} whose strength is considered by the warping coefficient $\lambda$, and the last term represents the spin-orbit interaction having $\alpha$ a  spin-orbit coupling constant, $E_0$ is a constant term, and $k_{\pm} = k_x \pm i k_y$ where $k_x$ ($k_y$) is the component of the wave vector along $\hat{x}$ ($\hat{y}$) direction. To obtain the Hamiltonian for the edge states, we consider a finite size system which is placed in the $x-y$ plane and is defined between the boundaries as $-d/2 < y < d/2$ along the $\hat{x}$-direction. Due to the broken translational symmetry for the $\hat{y}$-direction, the wave vector $k_y$ needs to be replaced by an operator $-i\partial_y$ and the eigenvalue equation for the upper and lower block matrix separately using the Schrodinger equation $h_s(k_x,-i\partial_y) \Phi_{0}^{s} = E_{s}\Phi_{0}^{s}$ has been solved where $\Phi_{\kappa}^{s} = e^{ik_x x}e^{\kappa y} \psi_{s}$ is the edge wavefunction for the above Hamiltonian having $\kappa$ as real numbers, and $s=+1$ ($-1$) for the upper, $\uparrow$ (lower, $\downarrow$) block. In principle, the resulting calculation is too complex and it is difficult to obtain the analytical expressions. To proceed, we first solve the edge wavefunction in the limit $\lambda \rightarrow 0$ and project the total Hamiltonian onto these edge wavefunctions $\{ \psi_0^{\uparrow,\gamma}(r),\psi_0^{\downarrow,\gamma}(r)\}$ with $\gamma = \pm$. We find that the Hamiltonian yields the dispersion $\varepsilon_{\pm} = E_0 \pm \alpha k_x \sigma_z$. Then, we find the elements of the effective edge Hamiltonian along the $\hat{x}$-direction using $H_{\text{edge}}^{\gamma ss'} = \langle \tilde{\Psi}_{0}^{s,\gamma}(r) \vert h_s(k_x,-i\partial_y) \vert \tilde{\Psi}_{0}^{s',\beta}(r) \rangle$, where $ \tilde{\Psi}_{0}^{s,\gamma}(r)$ is the normalized edge wavefunction . Finding the diagonal and off-diagonal elements taking into account the spin-orbit coupling and finite warping coefficient, the effective edge Hamiltonian after rotating the Pauli matrices $\sigma_z \rightarrow \sigma_x$, and $\sigma_y \rightarrow \sigma_z$ takes the form
\begin{equation}\label{eqn:BiSeHamil}
    H_{\text{edge}}^{\gamma}(k_x) = E_0\mathcal{I} + D_0 k_x^2\mathcal{I} + \lambda k_x^3 \sigma_x + \alpha k_x \sigma_z.
\end{equation} 
The detailed calculation to construct this edge Hamiltonian is provided in the Supplementary part. Now on adding the Zeeman energy term, the magnetic field interacts with the edge electrons through the Zeeman effect which results the modulation of the net current. This scenario is equivalent to the case of shifting the wave vector $k$ by $k-E_Z/A$ which ensures that for the non-linear current to be finite, the term $\propto \lambda$ is indispensable (this is a descendant of the well-known \textit{warping} term in 2D systems). There will be no non-linear response for this magnetic field orientation if $\lambda = 0$: hence an experimental measurement of a current at twice the applied frequency would immediately indicate that $\lambda$ is finite. The edge energy dispersion for this model at the different Zeeman energy $E_Z$ is demonstrated in Fig.~\ref{fig:dispersion}. Interestingly, there is another case where the finite Zeeman energy along the $\sigma_z$ opens a gap between conduction and valence bands in the dispersion, resulting the increase in the net current. In the similar way, we have derived the Hamiltonian for WTe$_2$ and the details of the derivation of the edge Hamiltonian are given in the Supplementary part.
\begin{figure}[!t]
    \centering
    \includegraphics[width=7cm, height=5.5cm]{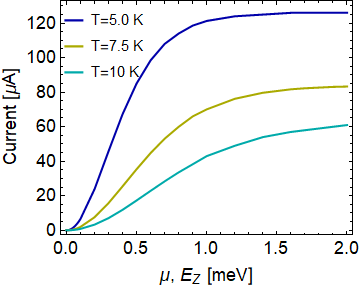}
    \caption{\textbf{Non-linear current as a function of the chemical potential}. Here the chemical potential ($\mu$) is set equal to the Zeeman energy ($E_Z$) at different low temperature values: 5, 7.5, and 10 K.}
    \label{fig:NL_BEmu}
\end{figure}

Firstly, we consider the case for the inversion symmetric topological insulator Bi$_2$Se$_3$ whose edge dispersion is given by Eq.~(\ref{eqn:BiSeHamil}). Now, on adding the Zeeman energy term $\propto \sigma_z$ the origin in the momentum space gets shifted and the dispersion here is described by $\varepsilon = \sqrt{(\lambda k-E_z/A)^6 + A^2k^2} \sigma_z$. We find that when $E_Z \rightarrow 0$, the second-order current vanishes as shown in Fig.~\ref{fig:NLR_main} due to the vanishing $f^d = f_0(\varepsilon^+) - f_0(\varepsilon^-)$ at $k=0$. However, the finite $E_Z$ yields the non-zero second-order current. Further, this current increases at low energy and then saturates at low temperatures when the Zeeman energy and the chemical potential become equal, as depicted in Fig.~\ref{fig:NL_BEmu}. But as the temperature rises $j^{(2)}$ decreases due to the broadening of the delta function. 

Secondly, for the WTe$_2$ model the edge dispersion is represented as $\varepsilon =  \varepsilon_0 \mathcal{I} + \varepsilon_3 \sigma_z$, where $\varepsilon_0 = D k^2$, and $\varepsilon_3 = \sqrt{(E_z + A k)^2 + (Ck+\lambda k^3)^2}$. Here the kinetic energy takes into account the particle hole asymmetry and the Zeeman term along $\hat{z}$-direction shifts the $k$ points by an amount $E_z/A$ as mentioned earlier. In the same way as with Bi$_2$Se$_3$, we find that a finite $\lambda$ leads to a finite value for the non-linear current. This can be probed by varying the Zeeman energy and has behavior depicted in Fig.~\ref{fig:NLR_main}. Moreover, in both cases the non-linearity can be increased by increasing the potential difference between the source and drain electrodes. 

Thirdly, for the edge states of quantum anomalous Hall insulators, the edge dispersion is $\varepsilon_0 = t k$, where $t$ is a material dependent parameter. This degenerate dispersion for two bands gives the non-linear current at the low temperature as $\delta(\mu)$ which survives only at zero chemical potential. It is to be noted that $\mu =0$ lies on the top of the valence band where the dispersion actually starts, thus in practice it will be rather challenging to observe the non-linear current for quantum anomalous Hall edge states. Finally, for the HgTe edge states we find that $j^{(2)}$ vanishes with the linear wave vector Hamiltonian \cite{durnev_PRB2016} along the $\hat{z}$-direction due to the absence of the warping and Zeeman field effects. The details for this system are provided in the Supplementary material. 

Interestingly, in the presence of the particle-hole asymmetry, one might expect the occurrence of the Zitterbewegung in the one dimensional non-linear current due to the exponential dependence of the special function $g(k)$. However, this does not happen in any situation here. We have seen that the non-linear current generates in the presence of finite Zeeman energy. If we take $E_Z = 0$, the function $g(0)$ comes out to be finite, but the difference in the distribution functions between two bands vanishes at the origin due to which the non-linear current becomes zero. On turning the Zeeman energy non-zero, the function $g(0)$ diverges due to the term $\varepsilon_3/v_0$, where $\varepsilon_3$ now acquires the Zeeman contribution. This also makes the Zitterbewegung term zero. Hence there is no Zitterbewegung occur in either way.

In summary, we have developed a general quantum kinetic theory for the non-linear transport in the ballistic regime and examine the Zitterbewegung terms. We test our theory for the edge states of the topological materials and find that the Kramers symmetry breaking term produce the finite non-linear current which further can be tuned by the Zeeman energy. In future, this theory can also be extended to study the non-linear transport of the Luttinger liquids. 
\section*{Methods}  We begin with the quantum Liouville equation for the density matrix $\rho$ within the ballistic limit in the eigen-state basis
\begin{equation}
   \hbar \pdv{\rho}{t} + i [H_0,\rho] + \hbar\bigg\{ v,\pdv{\rho}{x} \bigg\} = \mathcal{D}.
   \label{eqn:QLE}
\end{equation}
Here $H_0$ is the edge state Hamiltonian, $v = \hbar^{-1} (\partial H_0/\partial k - i [\mathcal{R}_{k}, H_0])$ is the velocity, where $\mathcal{R}_{k} = \langle u_{k} \vert \nabla \vert u_{k} \rangle$ is the Berry connection with $u_{k}$ an eigen vector corresponding to the model Hamiltonian of the system and $[\cdot,\cdot ]$, and $\{ \cdot, \cdot \}$ represent the commutator and anticommutator between two quantities respectively, the electric-field driving term $\mathcal{D} = -e\partial V/\partial x (\partial\rho / \partial k  + i [\mathcal{R}_{k},\rho])$, having $V$ the space dependent electrostatic potential. To find the solution of the differential equation, it is convenient to take the Fourier transform (FT) of the above equation with respect to space and time which reduces the Eq.~(\ref{eqn:QLE}) to
\begin{equation}
-i\hbar\omega \rho + i [H_0,\rho] - i q\hbar \{v, \rho\} = \mathcal{D}(q,\omega),
\label{eqn:FT}
\end{equation}
where the density matrix $\rho$ is a function of $(q,\omega)$. According to the definition of the driving term, $\mathcal{D}$ depends on the potential and the density matrix, and is independent of the time, thus the time FT of $\mathcal{D}$ becomes proportional to $\delta(\omega)$. This reduces the driving term $\mathcal{D}(q,\omega) = \delta(\omega) \mathcal{D}(q)$ which will contribute only at $\omega = 0$. Due to this reason, we neglect the frequency dependence nature of the solution for the density matrix. To stratify the resulting equation, we decompose the quantities in terms of $2 \times 2$ Pauli matrices, such as $\rho = S_0 \mathcal{I} + S_1 \sigma_x + S_2 \sigma_y + S_3 \sigma_z$, $v = v_0\mathcal{I} + v_1\sigma_x + v_2\sigma_y + v_3\sigma_z$ where $S_i$, and $v_i$ represent the components of the density matrix and the velocity associated with the corresponding Pauli matrices $\sigma_i$ respectively. However, the dispersion corresponding to the Hamiltonian of the system after diagonalizing the Hamiltonian can be written like $H = \varepsilon_0 \mathcal{I} + \varepsilon_3 \sigma_z$ having $\varepsilon_i$ the components of the full dispersion of the considered system. This decomposition leads to a set of four equations corresponding to the identity matrix and three Pauli matrices. The resulting equations can be naively expressed in the matrix form like $\mathcal{M} \rho = \mathcal{D}$, where $\mathcal{M}$ is the $4 \times 4$ matrix comprised of different components of the velocity and $q$, the wavevector, $\rho$, and $\mathcal{D}$ are four vector column matrices. The premultiplication of the inverse of the matrix $\mathcal{M}$ to the the matrix form and the inverse Fourier transform gives the solution of the density matrix in the real space which reads as 
\begin{equation}
    \rho(x) = \mathcal{M}^{-1}(x) \mathcal{D}(x).
\end{equation}
According to the basic definition, the electric current is the trace of the product of the velocity and the density matrix. Inserting the expressions for $v$, and $\rho$, one can find the current to order N in the potential 
\begin{equation}
\ba
j^{(N)} &\dps =  -\frac{e^2}{h} \int_{0}^{x} dx' \pd{V}{x'}p(x') + \frac{e^2}{h} \int_{d}^{x} dx' \pd{V}{x'}p(x').
\label{eqn:current}
\ea
\end{equation}
Here the $x$ integral is for the length of the channel where the potential is intact, $p(x') = [S_0^{(N-1)}(x')]_{k=0}$, where $S_0^{(N)}(x) = M_{11}^{-1}(x) \mathcal{D}_{0}^{(N)}(x) + M_{12}^{-1}(x) \mathcal{D}_{1}^{(N)}(x) + M_{13}^{-1}(x) \mathcal{D}_{2}^{(N)}(x) + M_{14}^{-1}(x) \mathcal{D}_{3}^{(N)}(x)$ and $d$ is the length of the channel. The direct contributions stemming from other elements of the density matrix such as $S_1$, $S_2$, and $S_3$ cancel out due to the cancellation of terms after taking the product of the matrix elements of the $\mathcal{M}^{-1}(x)$ with the velocity components, thus do not contribute to the main expression for the non-linear current. However, the indirect contributions of such terms still survive through the driving terms. The detailed derivation of Eq.~(\ref{eqn:current}) is given in the Supplementary material. Further, in Eq.~(\ref{eqn:current}), the first term is for the carriers moving towards the right side of the electrode or right movers and the second term for the left movers. For the linear order or $N=1$, the quantity $p(x)$ reduce to $S_0^{(0)}$ which is the equilibrium part of the density matrix and it is equivalent to $f_0^s$, where $f_0^s = f_0(\varepsilon^+) + f_0(\varepsilon^-)$ is the sum of the equilibrium distribution functions at the energies correspond to the conduction ($+$) and the valence ($-$) bands. The distribution function is independent of the space variable, thus can be pulled out from the integrand and the spatial integration over $x$ can be performed easily. The latter results the linear current as $e^2/h (V_L - V_R) f_0^s$. At the low temperature and when the chemical potential lies in one of the band, the current becomes $e^2/h (V_L - V_R)$. This is consistent with the Landauer-Buttiker formula for the conductance \cite{landauer_ZPB1987, datta_book}. However, if the chemical potential lies in between the gap of the valence and conduction bands, the current vanishes.

In the non-linear regime or the second-order case, the current becomes proportional to the first power of the density matrix $S_0^{(1)}$ at $k=0$. The latter quantity using the definitions for the driving terms can be written as
$S_0^{(1)} =\partial V/\partial x [ M_{11}^{-1}(x) \partial f_{0}^s/\partial k  + M_{12}^{-1}(x) \mathcal{R}_2f_{0}^d + M_{13}^{-1}(x) \mathcal{R}_1f_{0}^d + M_{14}^{-1}(x) \partial f_{0}^d/\partial k ]$, where $f_{0}^d$ refers the difference between the equilibrium distribution function at different bands. To solve the space integral, it is convenient to split the matrix elements into two parts $M_{ij}^{-1}(x) = [M_{ij}^{-1}]_a + [M_{ij}^{-1}]_b e^{-x g(k)}$, having the first part of the element as space independent while the other part depends. The forms of different elements of the inverse of the matrix $\mathcal{M}$ are mentioned in the Supplementary part. Here $g(k)$ is a function containing the combination of the velocity components which is defined as 
\begin{equation}
\ba
g(k) &\dps = \sqrt{\frac{(v_3^2 - v_0^2) \varepsilon_3^2}{\hbar^2 v_0^2 (v_0^2 - v_1^2 - v_2^2 - v_3^2)}}.
\label{eqn:gk}
\ea
\end{equation}
This term decides the Zitterbewegung or decaying nature of the current depending on the strengths of the velocity components. However, in the present study we find that at $k=0$ the terms associated with the exponential factor approaches to zero due to the diverging nature of $g(k)$ for model having finite value of the component of the velocity $v_0$. In other cases, this term does not emerge within the matrix elements. Thus we can drop this term for the further analysis. After this simplification, the non-linear current becomes
\begin{equation}
\ba
j^{(2)} &\dps = \frac{e^3(V_L - V_R)^2}{h} \bigg( M_{11}^{-1} \pdv{ f_{0}^s}{k} + M_{14}^{-1} \pdv{f_{0}^d}{k}  \bigg)_{k=0}.
\label{eqn:NLCmain}
\ea
\end{equation}
This is the general expression for the second-order current of the ballistic case.

\section*{Acknowledgments} This work was supported by the Australian Research Council Centre of Excellence in Future Low-Energy Electronics Technologies (project number CE170100039). PB acknowledges the National Key Research and Development Program of China (grant No. 2017YFA0303400), China postdoctoral science foundation (grant no. 2019M650461) and NSFC grant no. U1930402 for financial support.

\section*{data availability} Data sharing not applicable to this article as no data sets were generated or analyzed during this study.

\section*{Contributions} P.B., M.X.D. and D.C. performed the calculations. R.Q.W. and D.C. supervised the theoretical effort and L.W. designed the proposed architecture for experimental observation. P.B. and D.C. drafted the manuscript.

\section*{Competing Interests} The authors declare no competing interests.


\begin{thebibliography}{69}%
\makeatletter
\providecommand \@ifxundefined [1]{%
 \@ifx{#1\undefined}
}%
\providecommand \@ifnum [1]{%
 \ifnum #1\expandafter \@firstoftwo
 \else \expandafter \@secondoftwo
 \fi
}%
\providecommand \@ifx [1]{%
 \ifx #1\expandafter \@firstoftwo
 \else \expandafter \@secondoftwo
 \fi
}%
\providecommand \natexlab [1]{#1}%
\providecommand \enquote  [1]{``#1''}%
\providecommand \bibnamefont  [1]{#1}%
\providecommand \bibfnamefont [1]{#1}%
\providecommand \citenamefont [1]{#1}%
\providecommand \href@noop [0]{\@secondoftwo}%
\providecommand \href [0]{\begingroup \@sanitize@url \@href}%
\providecommand \@href[1]{\@@startlink{#1}\@@href}%
\providecommand \@@href[1]{\endgroup#1\@@endlink}%
\providecommand \@sanitize@url [0]{\catcode `\\12\catcode `\$12\catcode
  `\&12\catcode `\#12\catcode `\^12\catcode `\_12\catcode `\%12\relax}%
\providecommand \@@startlink[1]{}%
\providecommand \@@endlink[0]{}%
\providecommand \url  [0]{\begingroup\@sanitize@url \@url }%
\providecommand \@url [1]{\endgroup\@href {#1}{\urlprefix }}%
\providecommand \urlprefix  [0]{URL }%
\providecommand \Eprint [0]{\href }%
\providecommand \doibase [0]{https://doi.org/}%
\providecommand \selectlanguage [0]{\@gobble}%
\providecommand \bibinfo  [0]{\@secondoftwo}%
\providecommand \bibfield  [0]{\@secondoftwo}%
\providecommand \translation [1]{[#1]}%
\providecommand \BibitemOpen [0]{}%
\providecommand \bibitemStop [0]{}%
\providecommand \bibitemNoStop [0]{.\EOS\space}%
\providecommand \EOS [0]{\spacefactor3000\relax}%
\providecommand \BibitemShut  [1]{\csname bibitem#1\endcsname}%
\let\auto@bib@innerbib\@empty
\bibitem [{\citenamefont {Zhang}\ \emph {et~al.}(2009)\citenamefont {Zhang},
  \citenamefont {Liu}, \citenamefont {Qi}, \citenamefont {Dai}, \citenamefont
  {Fang},\ and\ \citenamefont {Zhang}}]{zhang_NP2009}%
  \BibitemOpen
  \bibfield  {author} {\bibinfo {author} {\bibfnamefont {H.}~\bibnamefont
  {Zhang}}, \bibinfo {author} {\bibfnamefont {C.-X.}\ \bibnamefont {Liu}},
  \bibinfo {author} {\bibfnamefont {X.-L.}\ \bibnamefont {Qi}}, \bibinfo
  {author} {\bibfnamefont {X.}~\bibnamefont {Dai}}, \bibinfo {author}
  {\bibfnamefont {Z.}~\bibnamefont {Fang}},\ and\ \bibinfo {author}
  {\bibfnamefont {S.-C.}\ \bibnamefont {Zhang}},\ }\bibfield  {title} {\bibinfo
  {title} {Topological insulators in \text{Bi}$_2$\text{Se}$_3$,
  \text{Bi}$_2$\text{Te}$_3$ and \text{Sb}$_2$\text{Te}$_3$ with a single dirac
  cone on the surface},\ }\href {https://doi.org/10.1038/nphys1270} {\bibfield
  {journal} {\bibinfo  {journal} {Nature Physics}\ }\textbf {\bibinfo {volume}
  {5}},\ \bibinfo {pages} {438} (\bibinfo {year} {2009})}\BibitemShut {NoStop}%
\bibitem [{\citenamefont {Hasan}\ and\ \citenamefont
  {Kane}(2010)}]{hasan_RMP2010}%
  \BibitemOpen
  \bibfield  {author} {\bibinfo {author} {\bibfnamefont {M.~Z.}\ \bibnamefont
  {Hasan}}\ and\ \bibinfo {author} {\bibfnamefont {C.~L.}\ \bibnamefont
  {Kane}},\ }\bibfield  {title} {\bibinfo {title} {Colloquium: Topological
  insulators},\ }\href {https://doi.org/10.1103/RevModPhys.82.3045} {\bibfield
  {journal} {\bibinfo  {journal} {Rev. Mod. Phys.}\ }\textbf {\bibinfo {volume}
  {82}},\ \bibinfo {pages} {3045} (\bibinfo {year} {2010})}\BibitemShut
  {NoStop}%
\bibitem [{\citenamefont {Liu}\ \emph {et~al.}(2010)\citenamefont {Liu},
  \citenamefont {Qi}, \citenamefont {Zhang}, \citenamefont {Dai}, \citenamefont
  {Fang},\ and\ \citenamefont {Zhang}}]{liu_PRB2010}%
  \BibitemOpen
  \bibfield  {author} {\bibinfo {author} {\bibfnamefont {C.-X.}\ \bibnamefont
  {Liu}}, \bibinfo {author} {\bibfnamefont {X.-L.}\ \bibnamefont {Qi}},
  \bibinfo {author} {\bibfnamefont {H.}~\bibnamefont {Zhang}}, \bibinfo
  {author} {\bibfnamefont {X.}~\bibnamefont {Dai}}, \bibinfo {author}
  {\bibfnamefont {Z.}~\bibnamefont {Fang}},\ and\ \bibinfo {author}
  {\bibfnamefont {S.-C.}\ \bibnamefont {Zhang}},\ }\bibfield  {title} {\bibinfo
  {title} {Model hamiltonian for topological insulators},\ }\href
  {https://doi.org/10.1103/PhysRevB.82.045122} {\bibfield  {journal} {\bibinfo
  {journal} {Phys. Rev. B}\ }\textbf {\bibinfo {volume} {82}},\ \bibinfo
  {pages} {045122} (\bibinfo {year} {2010})}\BibitemShut {NoStop}%
\bibitem [{\citenamefont {Chu}\ \emph {et~al.}(2011)\citenamefont {Chu},
  \citenamefont {Shan}, \citenamefont {Lu},\ and\ \citenamefont
  {Shen}}]{chu_PRB2011}%
  \BibitemOpen
  \bibfield  {author} {\bibinfo {author} {\bibfnamefont {R.-L.}\ \bibnamefont
  {Chu}}, \bibinfo {author} {\bibfnamefont {W.-Y.}\ \bibnamefont {Shan}},
  \bibinfo {author} {\bibfnamefont {J.}~\bibnamefont {Lu}},\ and\ \bibinfo
  {author} {\bibfnamefont {S.-Q.}\ \bibnamefont {Shen}},\ }\bibfield  {title}
  {\bibinfo {title} {Surface and edge states in topological semimetals},\
  }\href {https://doi.org/10.1103/PhysRevB.83.075110} {\bibfield  {journal}
  {\bibinfo  {journal} {Phys. Rev. B}\ }\textbf {\bibinfo {volume} {83}},\
  \bibinfo {pages} {075110} (\bibinfo {year} {2011})}\BibitemShut {NoStop}%
\bibitem [{\citenamefont {Adagideli}\ and\ \citenamefont
  {Bauer}(2005)}]{adagideli_PRL2005}%
  \BibitemOpen
  \bibfield  {author} {\bibinfo {author} {\bibfnamefont {I.}~\bibnamefont
  {Adagideli}}\ and\ \bibinfo {author} {\bibfnamefont {G.~E.~W.}\ \bibnamefont
  {Bauer}},\ }\bibfield  {title} {\bibinfo {title} {Intrinsic spin hall
  edges},\ }\href {https://doi.org/10.1103/PhysRevLett.95.256602} {\bibfield
  {journal} {\bibinfo  {journal} {Phys. Rev. Lett.}\ }\textbf {\bibinfo
  {volume} {95}},\ \bibinfo {pages} {256602} (\bibinfo {year}
  {2005})}\BibitemShut {NoStop}%
\bibitem [{\citenamefont {Yan}\ and\ \citenamefont
  {Zhang}(2012)}]{yan_RPP2012}%
  \BibitemOpen
  \bibfield  {author} {\bibinfo {author} {\bibfnamefont {B.}~\bibnamefont
  {Yan}}\ and\ \bibinfo {author} {\bibfnamefont {S.-C.}\ \bibnamefont
  {Zhang}},\ }\bibfield  {title} {\bibinfo {title} {Topological materials},\
  }\href {https://doi.org/10.1088/0034-4885/75/9/096501} {\bibfield  {journal}
  {\bibinfo  {journal} {Reports on Progress in Physics}\ }\textbf {\bibinfo
  {volume} {75}},\ \bibinfo {pages} {096501} (\bibinfo {year}
  {2012})}\BibitemShut {NoStop}%
\bibitem [{\citenamefont {Liu}\ \emph {et~al.}(2014)\citenamefont {Liu},
  \citenamefont {Zhou}, \citenamefont {Zhang}, \citenamefont {Wang},
  \citenamefont {Weng}, \citenamefont {Prabhakaran}, \citenamefont {Mo},
  \citenamefont {Shen}, \citenamefont {Fang}, \citenamefont {Dai},
  \citenamefont {Hussain},\ and\ \citenamefont {Chen}}]{Liu_Sc2014}%
  \BibitemOpen
  \bibfield  {author} {\bibinfo {author} {\bibfnamefont {Z.~K.}\ \bibnamefont
  {Liu}}, \bibinfo {author} {\bibfnamefont {B.}~\bibnamefont {Zhou}}, \bibinfo
  {author} {\bibfnamefont {Y.}~\bibnamefont {Zhang}}, \bibinfo {author}
  {\bibfnamefont {Z.~J.}\ \bibnamefont {Wang}}, \bibinfo {author}
  {\bibfnamefont {H.~M.}\ \bibnamefont {Weng}}, \bibinfo {author}
  {\bibfnamefont {D.}~\bibnamefont {Prabhakaran}}, \bibinfo {author}
  {\bibfnamefont {S.-K.}\ \bibnamefont {Mo}}, \bibinfo {author} {\bibfnamefont
  {Z.~X.}\ \bibnamefont {Shen}}, \bibinfo {author} {\bibfnamefont
  {Z.}~\bibnamefont {Fang}}, \bibinfo {author} {\bibfnamefont {X.}~\bibnamefont
  {Dai}}, \bibinfo {author} {\bibfnamefont {Z.}~\bibnamefont {Hussain}},\ and\
  \bibinfo {author} {\bibfnamefont {Y.~L.}\ \bibnamefont {Chen}},\ }\bibfield
  {title} {\bibinfo {title} {Discovery of a three-dimensional topological dirac
  semimetal, \text{Na}$_3$\text{Bi}},\ }\href
  {https://doi.org/10.1126/science.1245085} {\bibfield  {journal} {\bibinfo
  {journal} {Science}\ }\textbf {\bibinfo {volume} {343}},\ \bibinfo {pages}
  {864} (\bibinfo {year} {2014})}\BibitemShut {NoStop}%
\bibitem [{\citenamefont {Weber}\ \emph {et~al.}(2014)\citenamefont {Weber},
  \citenamefont {Tan}, \citenamefont {Mahapatra}, \citenamefont {Watson},
  \citenamefont {Ryu}, \citenamefont {Rahman}, \citenamefont {Hollenberg},
  \citenamefont {Klimeck},\ and\ \citenamefont {Simmons}}]{weber_NN2014}%
  \BibitemOpen
  \bibfield  {author} {\bibinfo {author} {\bibfnamefont {B.}~\bibnamefont
  {Weber}}, \bibinfo {author} {\bibfnamefont {Y.~H.~M.}\ \bibnamefont {Tan}},
  \bibinfo {author} {\bibfnamefont {S.}~\bibnamefont {Mahapatra}}, \bibinfo
  {author} {\bibfnamefont {T.~F.}\ \bibnamefont {Watson}}, \bibinfo {author}
  {\bibfnamefont {H.}~\bibnamefont {Ryu}}, \bibinfo {author} {\bibfnamefont
  {R.}~\bibnamefont {Rahman}}, \bibinfo {author} {\bibfnamefont {L.~C.~L.}\
  \bibnamefont {Hollenberg}}, \bibinfo {author} {\bibfnamefont
  {G.}~\bibnamefont {Klimeck}},\ and\ \bibinfo {author} {\bibfnamefont {M.~Y.}\
  \bibnamefont {Simmons}},\ }\bibfield  {title} {\bibinfo {title} {Spin
  blockade and exchange in coulomb-confined silicon double quantum dots},\
  }\href {https://doi.org/10.1038/nnano.2014.63} {\bibfield  {journal}
  {\bibinfo  {journal} {Nature Nanotechnology}\ }\textbf {\bibinfo {volume}
  {9}},\ \bibinfo {pages} {430} (\bibinfo {year} {2014})}\BibitemShut {NoStop}%
\bibitem [{\citenamefont {Farrell}\ and\ \citenamefont
  {Pereg-Barnea}(2016)}]{farrell_PRB2016}%
  \BibitemOpen
  \bibfield  {author} {\bibinfo {author} {\bibfnamefont {A.}~\bibnamefont
  {Farrell}}\ and\ \bibinfo {author} {\bibfnamefont {T.}~\bibnamefont
  {Pereg-Barnea}},\ }\bibfield  {title} {\bibinfo {title} {Edge-state transport
  in floquet topological insulators},\ }\href
  {https://doi.org/10.1103/PhysRevB.93.045121} {\bibfield  {journal} {\bibinfo
  {journal} {Phys. Rev. B}\ }\textbf {\bibinfo {volume} {93}},\ \bibinfo
  {pages} {045121} (\bibinfo {year} {2016})}\BibitemShut {NoStop}%
\bibitem [{\citenamefont {Yan}\ and\ \citenamefont
  {Felser}(2017)}]{yan_ARCMP2017}%
  \BibitemOpen
  \bibfield  {author} {\bibinfo {author} {\bibfnamefont {B.}~\bibnamefont
  {Yan}}\ and\ \bibinfo {author} {\bibfnamefont {C.}~\bibnamefont {Felser}},\
  }\bibfield  {title} {\bibinfo {title} {Topological materials: Weyl
  semimetals},\ }\href
  {https://doi.org/10.1146/annurev-conmatphys-031016-025458} {\bibfield
  {journal} {\bibinfo  {journal} {Annual Review of Condensed Matter Physics}\
  }\textbf {\bibinfo {volume} {8}},\ \bibinfo {pages} {337} (\bibinfo {year}
  {2017})}\BibitemShut {NoStop}%
\bibitem [{\citenamefont {Wang}\ and\ \citenamefont
  {Zhang}(2017)}]{wang_NM2017}%
  \BibitemOpen
  \bibfield  {author} {\bibinfo {author} {\bibfnamefont {J.}~\bibnamefont
  {Wang}}\ and\ \bibinfo {author} {\bibfnamefont {S.-C.}\ \bibnamefont
  {Zhang}},\ }\bibfield  {title} {\bibinfo {title} {Topological states of
  condensed matter},\ }\href {https://doi.org/10.1038/nmat5012} {\bibfield
  {journal} {\bibinfo  {journal} {Nature Materials}\ }\textbf {\bibinfo
  {volume} {16}},\ \bibinfo {pages} {1062} (\bibinfo {year}
  {2017})}\BibitemShut {NoStop}%
\bibitem [{\citenamefont {He}\ \emph {et~al.}(2018)\citenamefont {He},
  \citenamefont {Wang},\ and\ \citenamefont {Xue}}]{he_ARCM2018}%
  \BibitemOpen
  \bibfield  {author} {\bibinfo {author} {\bibfnamefont {K.}~\bibnamefont
  {He}}, \bibinfo {author} {\bibfnamefont {Y.}~\bibnamefont {Wang}},\ and\
  \bibinfo {author} {\bibfnamefont {Q.-K.}\ \bibnamefont {Xue}},\ }\bibfield
  {title} {\bibinfo {title} {Topological materials: Quantum anomalous hall
  system},\ }\href {https://doi.org/10.1146/annurev-conmatphys-033117-054144}
  {\bibfield  {journal} {\bibinfo  {journal} {Annual Review of Condensed Matter
  Physics}\ }\textbf {\bibinfo {volume} {9}},\ \bibinfo {pages} {329} (\bibinfo
  {year} {2018})}\BibitemShut {NoStop}%
\bibitem [{\citenamefont {Shen}(2017)}]{shen_book}%
  \BibitemOpen
  \bibfield  {author} {\bibinfo {author} {\bibfnamefont {S.-Q.}\ \bibnamefont
  {Shen}},\ }\bibinfo {title} {Topological dirac and weyl semimetals},\ in\
  \href {https://doi.org/10.1007/978-981-10-4606-3_11} {\emph {\bibinfo
  {booktitle} {Topological Insulators: Dirac Equation in Condensed Matter}}}\
  (\bibinfo  {publisher} {Springer Singapore},\ \bibinfo {address}
  {Singapore},\ \bibinfo {year} {2017})\ pp.\ \bibinfo {pages}
  {207--229}\BibitemShut {NoStop}%
\bibitem [{\citenamefont {Wang}\ \emph {et~al.}(2016)\citenamefont {Wang},
  \citenamefont {Wang}, \citenamefont {Lu}, \citenamefont {Jiang},\ and\
  \citenamefont {Yang}}]{wang_Nat2016}%
  \BibitemOpen
  \bibfield  {author} {\bibinfo {author} {\bibfnamefont {Y.}~\bibnamefont
  {Wang}}, \bibinfo {author} {\bibfnamefont {S.-S.}\ \bibnamefont {Wang}},
  \bibinfo {author} {\bibfnamefont {Y.}~\bibnamefont {Lu}}, \bibinfo {author}
  {\bibfnamefont {J.}~\bibnamefont {Jiang}},\ and\ \bibinfo {author}
  {\bibfnamefont {S.~A.}\ \bibnamefont {Yang}},\ }\bibfield  {title} {\bibinfo
  {title} {Strain-induced isostructural and magnetic phase transitions in
  monolayer \text{MoN}$_2$},\ }\href
  {https://doi.org/10.1021/acs.nanolett.6b01841} {\bibfield  {journal}
  {\bibinfo  {journal} {Nano Letters}\ }\textbf {\bibinfo {volume} {16}},\
  \bibinfo {pages} {4576} (\bibinfo {year} {2016})}\BibitemShut {NoStop}%
\bibitem [{\citenamefont {Collins}\ \emph {et~al.}(2018)\citenamefont
  {Collins}, \citenamefont {Tadich}, \citenamefont {Wu}, \citenamefont {Gomes},
  \citenamefont {Rodrigues}, \citenamefont {Liu}, \citenamefont {Hellerstedt},
  \citenamefont {Ryu}, \citenamefont {Tang}, \citenamefont {Mo}, \citenamefont
  {Adam}, \citenamefont {Yang}, \citenamefont {Fuhrer},\ and\ \citenamefont
  {Edmonds}}]{collins_NP2018}%
  \BibitemOpen
  \bibfield  {author} {\bibinfo {author} {\bibfnamefont {J.~L.}\ \bibnamefont
  {Collins}}, \bibinfo {author} {\bibfnamefont {A.}~\bibnamefont {Tadich}},
  \bibinfo {author} {\bibfnamefont {W.}~\bibnamefont {Wu}}, \bibinfo {author}
  {\bibfnamefont {L.~C.}\ \bibnamefont {Gomes}}, \bibinfo {author}
  {\bibfnamefont {J.~N.~B.}\ \bibnamefont {Rodrigues}}, \bibinfo {author}
  {\bibfnamefont {C.}~\bibnamefont {Liu}}, \bibinfo {author} {\bibfnamefont
  {J.}~\bibnamefont {Hellerstedt}}, \bibinfo {author} {\bibfnamefont
  {H.}~\bibnamefont {Ryu}}, \bibinfo {author} {\bibfnamefont {S.}~\bibnamefont
  {Tang}}, \bibinfo {author} {\bibfnamefont {S.-K.}\ \bibnamefont {Mo}},
  \bibinfo {author} {\bibfnamefont {S.}~\bibnamefont {Adam}}, \bibinfo {author}
  {\bibfnamefont {S.~A.}\ \bibnamefont {Yang}}, \bibinfo {author}
  {\bibfnamefont {M.~S.}\ \bibnamefont {Fuhrer}},\ and\ \bibinfo {author}
  {\bibfnamefont {M.~T.}\ \bibnamefont {Edmonds}},\ }\bibfield  {title}
  {\bibinfo {title} {Electric-field-tuned topological phase transition in
  ultrathin na3bi},\ }\href {https://doi.org/10.1038/s41586-018-0788-5}
  {\bibfield  {journal} {\bibinfo  {journal} {Nature}\ }\textbf {\bibinfo
  {volume} {564}},\ \bibinfo {pages} {390} (\bibinfo {year}
  {2018})}\BibitemShut {NoStop}%
\bibitem [{\citenamefont {Scharf}\ \emph {et~al.}(2015)\citenamefont {Scharf},
  \citenamefont {Matos-Abiague}, \citenamefont {\ifmmode \check{Z}\else
  \v{Z}\fi{}uti\ifmmode~\acute{c}\else \'{c}\fi{}},\ and\ \citenamefont
  {Fabian}}]{scharf_PRB2015}%
  \BibitemOpen
  \bibfield  {author} {\bibinfo {author} {\bibfnamefont {B.}~\bibnamefont
  {Scharf}}, \bibinfo {author} {\bibfnamefont {A.}~\bibnamefont
  {Matos-Abiague}}, \bibinfo {author} {\bibfnamefont {I.}~\bibnamefont
  {\ifmmode \check{Z}\else \v{Z}\fi{}uti\ifmmode~\acute{c}\else \'{c}\fi{}}},\
  and\ \bibinfo {author} {\bibfnamefont {J.}~\bibnamefont {Fabian}},\
  }\bibfield  {title} {\bibinfo {title} {Probing topological transitions in
  \text{HgTe}/\text{CdTe} quantum wells by magneto-optical measurements},\
  }\href {https://doi.org/10.1103/PhysRevB.91.235433} {\bibfield  {journal}
  {\bibinfo  {journal} {Phys. Rev. B}\ }\textbf {\bibinfo {volume} {91}},\
  \bibinfo {pages} {235433} (\bibinfo {year} {2015})}\BibitemShut {NoStop}%
\bibitem [{\citenamefont {Rostami}\ and\ \citenamefont {Juri\ifmmode
  \check{c}\else \v{c}\fi{}i\ifmmode~\acute{c}\else
  \'{c}\fi{}}(2020)}]{rostami_PRR2020}%
  \BibitemOpen
  \bibfield  {author} {\bibinfo {author} {\bibfnamefont {H.}~\bibnamefont
  {Rostami}}\ and\ \bibinfo {author} {\bibfnamefont {V.}~\bibnamefont
  {Juri\ifmmode \check{c}\else \v{c}\fi{}i\ifmmode~\acute{c}\else
  \'{c}\fi{}}},\ }\bibfield  {title} {\bibinfo {title} {Probing quantum
  criticality using nonlinear hall effect in a metallic dirac system},\ }\href
  {https://doi.org/10.1103/PhysRevResearch.2.013069} {\bibfield  {journal}
  {\bibinfo  {journal} {Phys. Rev. Research}\ }\textbf {\bibinfo {volume}
  {2}},\ \bibinfo {pages} {013069} (\bibinfo {year} {2020})}\BibitemShut
  {NoStop}%
\bibitem [{\citenamefont {Nemnes}\ \emph {et~al.}(2004)\citenamefont {Nemnes},
  \citenamefont {Wulf},\ and\ \citenamefont {Racec}}]{nemnes_JAP2004}%
  \BibitemOpen
  \bibfield  {author} {\bibinfo {author} {\bibfnamefont {G.~A.}\ \bibnamefont
  {Nemnes}}, \bibinfo {author} {\bibfnamefont {U.}~\bibnamefont {Wulf}},\ and\
  \bibinfo {author} {\bibfnamefont {P.~N.}\ \bibnamefont {Racec}},\ }\bibfield
  {title} {\bibinfo {title} {Nano-transistors in the landauer–büttiker
  formalism},\ }\href {https://doi.org/10.1063/1.1748858} {\bibfield  {journal}
  {\bibinfo  {journal} {Journal of Applied Physics}\ }\textbf {\bibinfo
  {volume} {96}},\ \bibinfo {pages} {596} (\bibinfo {year} {2004})}\BibitemShut
  {NoStop}%
\bibitem [{\citenamefont {Fabian}\ and\ \citenamefont {\ifmmode \check{Z}\else
  \v{Z}\fi{}uti\ifmmode~\acute{c}\else \'{c}\fi{}}(2004)}]{fabian_PRB2004}%
  \BibitemOpen
  \bibfield  {author} {\bibinfo {author} {\bibfnamefont {J.}~\bibnamefont
  {Fabian}}\ and\ \bibinfo {author} {\bibfnamefont {I.}~\bibnamefont {\ifmmode
  \check{Z}\else \v{Z}\fi{}uti\ifmmode~\acute{c}\else \'{c}\fi{}}},\ }\bibfield
   {title} {\bibinfo {title} {Spin-polarized current amplification and spin
  injection in magnetic bipolar transistors},\ }\href
  {https://doi.org/10.1103/PhysRevB.69.115314} {\bibfield  {journal} {\bibinfo
  {journal} {Phys. Rev. B}\ }\textbf {\bibinfo {volume} {69}},\ \bibinfo
  {pages} {115314} (\bibinfo {year} {2004})}\BibitemShut {NoStop}%
\bibitem [{\citenamefont {Huang}\ \emph {et~al.}(2007)\citenamefont {Huang},
  \citenamefont {Monsma},\ and\ \citenamefont {Appelbaum}}]{huang_APL2007}%
  \BibitemOpen
  \bibfield  {author} {\bibinfo {author} {\bibfnamefont {B.}~\bibnamefont
  {Huang}}, \bibinfo {author} {\bibfnamefont {D.~J.}\ \bibnamefont {Monsma}},\
  and\ \bibinfo {author} {\bibfnamefont {I.}~\bibnamefont {Appelbaum}},\
  }\bibfield  {title} {\bibinfo {title} {Experimental realization of a silicon
  spin field-effect transistor},\ }\href {https://doi.org/10.1063/1.2770656}
  {\bibfield  {journal} {\bibinfo  {journal} {Applied Physics Letters}\
  }\textbf {\bibinfo {volume} {91}},\ \bibinfo {pages} {072501} (\bibinfo
  {year} {2007})}\BibitemShut {NoStop}%
\bibitem [{\citenamefont {Wray}(2012)}]{wray_Nat2012}%
  \BibitemOpen
  \bibfield  {author} {\bibinfo {author} {\bibfnamefont {L.~A.}\ \bibnamefont
  {Wray}},\ }\bibfield  {title} {\bibinfo {title} {Topological transistor},\
  }\href {https://doi.org/10.1038/nphys2410} {\bibfield  {journal} {\bibinfo
  {journal} {Nature Physics}\ }\textbf {\bibinfo {volume} {8}},\ \bibinfo
  {pages} {705} (\bibinfo {year} {2012})}\BibitemShut {NoStop}%
\bibitem [{\citenamefont {Vicarelli}\ \emph {et~al.}(2012)\citenamefont
  {Vicarelli}, \citenamefont {Vitiello}, \citenamefont {Coquillat},
  \citenamefont {Lombardo}, \citenamefont {Ferrari}, \citenamefont {Knap},
  \citenamefont {Polini}, \citenamefont {Pellegrini},\ and\ \citenamefont
  {Tredicucci}}]{vicarelli_NM2012}%
  \BibitemOpen
  \bibfield  {author} {\bibinfo {author} {\bibfnamefont {L.}~\bibnamefont
  {Vicarelli}}, \bibinfo {author} {\bibfnamefont {M.~S.}\ \bibnamefont
  {Vitiello}}, \bibinfo {author} {\bibfnamefont {D.}~\bibnamefont {Coquillat}},
  \bibinfo {author} {\bibfnamefont {A.}~\bibnamefont {Lombardo}}, \bibinfo
  {author} {\bibfnamefont {A.~C.}\ \bibnamefont {Ferrari}}, \bibinfo {author}
  {\bibfnamefont {W.}~\bibnamefont {Knap}}, \bibinfo {author} {\bibfnamefont
  {M.}~\bibnamefont {Polini}}, \bibinfo {author} {\bibfnamefont
  {V.}~\bibnamefont {Pellegrini}},\ and\ \bibinfo {author} {\bibfnamefont
  {A.}~\bibnamefont {Tredicucci}},\ }\bibfield  {title} {\bibinfo {title}
  {Graphene field-effect transistors as room-temperature terahertz detectors},\
  }\href {https://doi.org/10.1038/nmat3417} {\bibfield  {journal} {\bibinfo
  {journal} {Nature Materials}\ }\textbf {\bibinfo {volume} {11}},\ \bibinfo
  {pages} {865} (\bibinfo {year} {2012})}\BibitemShut {NoStop}%
\bibitem [{\citenamefont {Cheng}\ \emph {et~al.}(2016)\citenamefont {Cheng},
  \citenamefont {Daniels}, \citenamefont {Zhu},\ and\ \citenamefont
  {Xiao}}]{cheng_SR2016}%
  \BibitemOpen
  \bibfield  {author} {\bibinfo {author} {\bibfnamefont {R.}~\bibnamefont
  {Cheng}}, \bibinfo {author} {\bibfnamefont {M.~W.}\ \bibnamefont {Daniels}},
  \bibinfo {author} {\bibfnamefont {J.-G.}\ \bibnamefont {Zhu}},\ and\ \bibinfo
  {author} {\bibfnamefont {D.}~\bibnamefont {Xiao}},\ }\bibfield  {title}
  {\bibinfo {title} {Antiferromagnetic spin wave field-effect transistor},\
  }\href {https://doi.org/10.1038/srep24223} {\bibfield  {journal} {\bibinfo
  {journal} {Scientific Reports}\ }\textbf {\bibinfo {volume} {6}},\ \bibinfo
  {pages} {24223} (\bibinfo {year} {2016})}\BibitemShut {NoStop}%
\bibitem [{\citenamefont {Landauer}(1987)}]{landauer_ZPB1987}%
  \BibitemOpen
  \bibfield  {author} {\bibinfo {author} {\bibfnamefont {R.}~\bibnamefont
  {Landauer}},\ }\bibfield  {title} {\bibinfo {title} {Electrical transport in
  open and closed systems},\ }\href {https://doi.org/10.1007/BF01304229}
  {\bibfield  {journal} {\bibinfo  {journal} {Zeitschrift für Physik B
  Condensed Matter}\ }\textbf {\bibinfo {volume} {68}},\ \bibinfo {pages} {217}
  (\bibinfo {year} {1987})}\BibitemShut {NoStop}%
\bibitem [{\citenamefont {Larkin}\ and\ \citenamefont
  {Khmel\'nitskii}(1986)}]{larkin_JETP1986}%
  \BibitemOpen
  \bibfield  {author} {\bibinfo {author} {\bibfnamefont {A.~I.}\ \bibnamefont
  {Larkin}}\ and\ \bibinfo {author} {\bibfnamefont {D.~E.}\ \bibnamefont
  {Khmel\'nitskii}},\ }\bibfield  {title} {\bibinfo {title} {Mesoscopic
  fluctuations of current-voltage characteristics},\ }\href@noop {} {\bibfield
  {journal} {\bibinfo  {journal} {JETP}\ }\textbf {\bibinfo {volume} {64}},\
  \bibinfo {pages} {1815} (\bibinfo {year} {1986})}\BibitemShut {NoStop}%
\bibitem [{\citenamefont {Bagwell}\ and\ \citenamefont
  {Orlando}(1989)}]{bagwell_PRB1989}%
  \BibitemOpen
  \bibfield  {author} {\bibinfo {author} {\bibfnamefont {P.~F.}\ \bibnamefont
  {Bagwell}}\ and\ \bibinfo {author} {\bibfnamefont {T.~P.}\ \bibnamefont
  {Orlando}},\ }\bibfield  {title} {\bibinfo {title} {Landauer's conductance
  formula and its generalization to finite voltages},\ }\href
  {https://doi.org/10.1103/PhysRevB.40.1456} {\bibfield  {journal} {\bibinfo
  {journal} {Phys. Rev. B}\ }\textbf {\bibinfo {volume} {40}},\ \bibinfo
  {pages} {1456} (\bibinfo {year} {1989})}\BibitemShut {NoStop}%
\bibitem [{\citenamefont {Meir}\ and\ \citenamefont
  {Wingreen}(1992)}]{meir_PRL1992}%
  \BibitemOpen
  \bibfield  {author} {\bibinfo {author} {\bibfnamefont {Y.}~\bibnamefont
  {Meir}}\ and\ \bibinfo {author} {\bibfnamefont {N.~S.}\ \bibnamefont
  {Wingreen}},\ }\bibfield  {title} {\bibinfo {title} {Landauer formula for the
  current through an interacting electron region},\ }\href
  {https://doi.org/10.1103/PhysRevLett.68.2512} {\bibfield  {journal} {\bibinfo
   {journal} {Phys. Rev. Lett.}\ }\textbf {\bibinfo {volume} {68}},\ \bibinfo
  {pages} {2512} (\bibinfo {year} {1992})}\BibitemShut {NoStop}%
\bibitem [{\citenamefont {Landauer}(1992)}]{landauer_PhyScr1992}%
  \BibitemOpen
  \bibfield  {author} {\bibinfo {author} {\bibfnamefont {R.}~\bibnamefont
  {Landauer}},\ }\bibfield  {title} {\bibinfo {title} {Conductance from
  transmission: common sense points},\ }\href
  {https://doi.org/10.1088/0031-8949/1992/t42/020} {\bibfield  {journal}
  {\bibinfo  {journal} {Physica Scripta}\ }\textbf {\bibinfo {volume} {T42}},\
  \bibinfo {pages} {110} (\bibinfo {year} {1992})}\BibitemShut {NoStop}%
\bibitem [{\citenamefont {Datta}(1993)}]{datta_book1993}%
  \BibitemOpen
  \bibfield  {author} {\bibinfo {author} {\bibfnamefont {S.}~\bibnamefont
  {Datta}},\ }\href {https://doi.org/10.1007/978-1-4899-2415-5_8} {\emph
  {\bibinfo {title} {Physics of Low-Dimensional Semiconductor Structures}}},\
  edited by\ \bibinfo {editor} {\bibfnamefont {P.}~\bibnamefont {Butcher}}\
  and\ \bibinfo {editor} {\bibfnamefont {M.~P.}\ \bibnamefont {March},
  \bibfnamefont {Norman H.~Tosi}}\ (\bibinfo  {publisher} {Springer US},\
  \bibinfo {address} {Boston, MA},\ \bibinfo {year} {1993})\ pp.\ \bibinfo
  {pages} {299--331}\BibitemShut {NoStop}%
\bibitem [{\citenamefont {Datta}(2005)}]{datta_book}%
  \BibitemOpen
  \bibfield  {author} {\bibinfo {author} {\bibfnamefont {S.}~\bibnamefont
  {Datta}},\ }\href {https://doi.org/10.1017/CBO9781139164313} {\emph {\bibinfo
  {title} {Quantum Transport: Atom to Transistor}}}\ (\bibinfo  {publisher}
  {Cambridge University Press},\ \bibinfo {year} {2005})\BibitemShut {NoStop}%
\bibitem [{\citenamefont {Green}\ and\ \citenamefont
  {Das}(2000)}]{green_JPCM2000}%
  \BibitemOpen
  \bibfield  {author} {\bibinfo {author} {\bibfnamefont {F.}~\bibnamefont
  {Green}}\ and\ \bibinfo {author} {\bibfnamefont {M.~P.}\ \bibnamefont
  {Das}},\ }\bibfield  {title} {\bibinfo {title} {Coulomb screening in
  mesoscopic noise: a kinetic approach},\ }\href
  {https://doi.org/10.1088/0953-8984/12/24/315} {\bibfield  {journal} {\bibinfo
   {journal} {Journal of Physics: Condensed Matter}\ }\textbf {\bibinfo
  {volume} {12}},\ \bibinfo {pages} {5251} (\bibinfo {year}
  {2000})}\BibitemShut {NoStop}%
\bibitem [{\citenamefont {Liang}\ \emph {et~al.}(2001)\citenamefont {Liang},
  \citenamefont {Bockrath}, \citenamefont {Bozovic}, \citenamefont {Hafner},
  \citenamefont {Tinkham},\ and\ \citenamefont {Park}}]{liang_Nat2001}%
  \BibitemOpen
  \bibfield  {author} {\bibinfo {author} {\bibfnamefont {W.}~\bibnamefont
  {Liang}}, \bibinfo {author} {\bibfnamefont {M.}~\bibnamefont {Bockrath}},
  \bibinfo {author} {\bibfnamefont {D.}~\bibnamefont {Bozovic}}, \bibinfo
  {author} {\bibfnamefont {J.~H.}\ \bibnamefont {Hafner}}, \bibinfo {author}
  {\bibfnamefont {M.}~\bibnamefont {Tinkham}},\ and\ \bibinfo {author}
  {\bibfnamefont {H.}~\bibnamefont {Park}},\ }\bibfield  {title} {\bibinfo
  {title} {Fabry - perot interference in a nanotube electron waveguide},\
  }\href {https://doi.org/10.1038/35079517} {\bibfield  {journal} {\bibinfo
  {journal} {Nature}\ }\textbf {\bibinfo {volume} {411}},\ \bibinfo {pages}
  {665} (\bibinfo {year} {2001})}\BibitemShut {NoStop}%
\bibitem [{\citenamefont {Jeong}\ \emph {et~al.}(2010)\citenamefont {Jeong},
  \citenamefont {Kim}, \citenamefont {Luisier}, \citenamefont {Datta},\ and\
  \citenamefont {Lundstrom}}]{Jeong_JAP2010}%
  \BibitemOpen
  \bibfield  {author} {\bibinfo {author} {\bibfnamefont {C.}~\bibnamefont
  {Jeong}}, \bibinfo {author} {\bibfnamefont {R.}~\bibnamefont {Kim}}, \bibinfo
  {author} {\bibfnamefont {M.}~\bibnamefont {Luisier}}, \bibinfo {author}
  {\bibfnamefont {S.}~\bibnamefont {Datta}},\ and\ \bibinfo {author}
  {\bibfnamefont {M.}~\bibnamefont {Lundstrom}},\ }\bibfield  {title} {\bibinfo
  {title} {On landauer versus boltzmann and full band versus effective mass
  evaluation of thermoelectric transport coefficients},\ }\href
  {https://doi.org/10.1063/1.3291120} {\bibfield  {journal} {\bibinfo
  {journal} {Journal of Applied Physics}\ }\textbf {\bibinfo {volume} {107}},\
  \bibinfo {pages} {023707} (\bibinfo {year} {2010})}\BibitemShut {NoStop}%
\bibitem [{\citenamefont {Knez}\ \emph {et~al.}(2011)\citenamefont {Knez},
  \citenamefont {Du},\ and\ \citenamefont {Sullivan}}]{knez_PRL2011}%
  \BibitemOpen
  \bibfield  {author} {\bibinfo {author} {\bibfnamefont {I.}~\bibnamefont
  {Knez}}, \bibinfo {author} {\bibfnamefont {R.-R.}\ \bibnamefont {Du}},\ and\
  \bibinfo {author} {\bibfnamefont {G.}~\bibnamefont {Sullivan}},\ }\bibfield
  {title} {\bibinfo {title} {Evidence for helical edge modes in inverted
  $\mathrm{InAs}/\mathrm{GaSb}$ quantum wells},\ }\href
  {https://doi.org/10.1103/PhysRevLett.107.136603} {\bibfield  {journal}
  {\bibinfo  {journal} {Phys. Rev. Lett.}\ }\textbf {\bibinfo {volume} {107}},\
  \bibinfo {pages} {136603} (\bibinfo {year} {2011})}\BibitemShut {NoStop}%
\bibitem [{\citenamefont {Cheianov}\ and\ \citenamefont
  {Glazman}(2013)}]{vadim_PRL2013}%
  \BibitemOpen
  \bibfield  {author} {\bibinfo {author} {\bibfnamefont {V.}~\bibnamefont
  {Cheianov}}\ and\ \bibinfo {author} {\bibfnamefont {L.~I.}\ \bibnamefont
  {Glazman}},\ }\bibfield  {title} {\bibinfo {title} {Mesoscopic fluctuations
  of conductance of a helical edge contaminated by magnetic impurities},\
  }\href {https://doi.org/10.1103/PhysRevLett.110.206803} {\bibfield  {journal}
  {\bibinfo  {journal} {Phys. Rev. Lett.}\ }\textbf {\bibinfo {volume} {110}},\
  \bibinfo {pages} {206803} (\bibinfo {year} {2013})}\BibitemShut {NoStop}%
\bibitem [{\citenamefont {Wang}\ and\ \citenamefont
  {Kruglyak}(2014)}]{wang_JNS2014}%
  \BibitemOpen
  \bibfield  {author} {\bibinfo {author} {\bibfnamefont {X.}~\bibnamefont
  {Wang}}\ and\ \bibinfo {author} {\bibfnamefont {Y.}~\bibnamefont
  {Kruglyak}},\ }\bibfield  {title} {\bibinfo {title} {Landauer-datta-lundstrom
  generalized transport model for nanoelectronics},\ }\href
  {https://doi.org/10.1155/2014/725420} {\bibfield  {journal} {\bibinfo
  {journal} {Journal of Nanoscience}\ }\textbf {\bibinfo {volume} {2014}},\
  \bibinfo {pages} {725420} (\bibinfo {year} {2014})}\BibitemShut {NoStop}%
\bibitem [{\citenamefont {Knez}\ \emph {et~al.}(2014)\citenamefont {Knez},
  \citenamefont {Rettner}, \citenamefont {Yang}, \citenamefont {Parkin},
  \citenamefont {Du}, \citenamefont {Du},\ and\ \citenamefont
  {Sullivan}}]{knez_PRL2014}%
  \BibitemOpen
  \bibfield  {author} {\bibinfo {author} {\bibfnamefont {I.}~\bibnamefont
  {Knez}}, \bibinfo {author} {\bibfnamefont {C.~T.}\ \bibnamefont {Rettner}},
  \bibinfo {author} {\bibfnamefont {S.-H.}\ \bibnamefont {Yang}}, \bibinfo
  {author} {\bibfnamefont {S.~S.~P.}\ \bibnamefont {Parkin}}, \bibinfo {author}
  {\bibfnamefont {L.}~\bibnamefont {Du}}, \bibinfo {author} {\bibfnamefont
  {R.-R.}\ \bibnamefont {Du}},\ and\ \bibinfo {author} {\bibfnamefont
  {G.}~\bibnamefont {Sullivan}},\ }\bibfield  {title} {\bibinfo {title}
  {Observation of edge transport in the disordered regime of topologically
  insulating $\mathrm{InAs}/\mathrm{GaSb}$ quantum wells},\ }\href
  {https://doi.org/10.1103/PhysRevLett.112.026602} {\bibfield  {journal}
  {\bibinfo  {journal} {Phys. Rev. Lett.}\ }\textbf {\bibinfo {volume} {112}},\
  \bibinfo {pages} {026602} (\bibinfo {year} {2014})}\BibitemShut {NoStop}%
\bibitem [{\citenamefont {Du}\ \emph {et~al.}(2015)\citenamefont {Du},
  \citenamefont {Knez}, \citenamefont {Sullivan},\ and\ \citenamefont
  {Du}}]{Du_PRL2015}%
  \BibitemOpen
  \bibfield  {author} {\bibinfo {author} {\bibfnamefont {L.}~\bibnamefont
  {Du}}, \bibinfo {author} {\bibfnamefont {I.}~\bibnamefont {Knez}}, \bibinfo
  {author} {\bibfnamefont {G.}~\bibnamefont {Sullivan}},\ and\ \bibinfo
  {author} {\bibfnamefont {R.-R.}\ \bibnamefont {Du}},\ }\bibfield  {title}
  {\bibinfo {title} {Robust helical edge transport in gated
  $\mathrm{InAs}/\mathrm{GaSb}$ bilayers},\ }\href
  {https://doi.org/10.1103/PhysRevLett.114.096802} {\bibfield  {journal}
  {\bibinfo  {journal} {Phys. Rev. Lett.}\ }\textbf {\bibinfo {volume} {114}},\
  \bibinfo {pages} {096802} (\bibinfo {year} {2015})}\BibitemShut {NoStop}%
\bibitem [{\citenamefont {Li}\ \emph {et~al.}(2015)\citenamefont {Li},
  \citenamefont {Wang}, \citenamefont {Fu}, \citenamefont {Du}, \citenamefont
  {Schreiber}, \citenamefont {Mu}, \citenamefont {Liu}, \citenamefont
  {Sullivan}, \citenamefont {Cs\'athy}, \citenamefont {Lin},\ and\
  \citenamefont {Du}}]{Li_PRL2015}%
  \BibitemOpen
  \bibfield  {author} {\bibinfo {author} {\bibfnamefont {T.}~\bibnamefont
  {Li}}, \bibinfo {author} {\bibfnamefont {P.}~\bibnamefont {Wang}}, \bibinfo
  {author} {\bibfnamefont {H.}~\bibnamefont {Fu}}, \bibinfo {author}
  {\bibfnamefont {L.}~\bibnamefont {Du}}, \bibinfo {author} {\bibfnamefont
  {K.~A.}\ \bibnamefont {Schreiber}}, \bibinfo {author} {\bibfnamefont
  {X.}~\bibnamefont {Mu}}, \bibinfo {author} {\bibfnamefont {X.}~\bibnamefont
  {Liu}}, \bibinfo {author} {\bibfnamefont {G.}~\bibnamefont {Sullivan}},
  \bibinfo {author} {\bibfnamefont {G.~A.}\ \bibnamefont {Cs\'athy}}, \bibinfo
  {author} {\bibfnamefont {X.}~\bibnamefont {Lin}},\ and\ \bibinfo {author}
  {\bibfnamefont {R.-R.}\ \bibnamefont {Du}},\ }\bibfield  {title} {\bibinfo
  {title} {Observation of a helical luttinger liquid in
  $\mathrm{InAs}/\mathrm{GaSb}$ quantum spin hall edges},\ }\href
  {https://doi.org/10.1103/PhysRevLett.115.136804} {\bibfield  {journal}
  {\bibinfo  {journal} {Phys. Rev. Lett.}\ }\textbf {\bibinfo {volume} {115}},\
  \bibinfo {pages} {136804} (\bibinfo {year} {2015})}\BibitemShut {NoStop}%
\bibitem [{\citenamefont {Fei}\ \emph {et~al.}(2017)\citenamefont {Fei},
  \citenamefont {Palomaki}, \citenamefont {Wu}, \citenamefont {Zhao},
  \citenamefont {Cai}, \citenamefont {Sun}, \citenamefont {Nguyen},
  \citenamefont {Finney}, \citenamefont {Xu},\ and\ \citenamefont
  {Cobden}}]{fei_NatPhy2017}%
  \BibitemOpen
  \bibfield  {author} {\bibinfo {author} {\bibfnamefont {Z.}~\bibnamefont
  {Fei}}, \bibinfo {author} {\bibfnamefont {T.}~\bibnamefont {Palomaki}},
  \bibinfo {author} {\bibfnamefont {S.}~\bibnamefont {Wu}}, \bibinfo {author}
  {\bibfnamefont {W.}~\bibnamefont {Zhao}}, \bibinfo {author} {\bibfnamefont
  {X.}~\bibnamefont {Cai}}, \bibinfo {author} {\bibfnamefont {B.}~\bibnamefont
  {Sun}}, \bibinfo {author} {\bibfnamefont {P.}~\bibnamefont {Nguyen}},
  \bibinfo {author} {\bibfnamefont {J.}~\bibnamefont {Finney}}, \bibinfo
  {author} {\bibfnamefont {X.}~\bibnamefont {Xu}},\ and\ \bibinfo {author}
  {\bibfnamefont {D.~H.}\ \bibnamefont {Cobden}},\ }\bibfield  {title}
  {\bibinfo {title} {Edge conduction in monolayer \text{WTe}$_2$},\ }\href
  {https://doi.org/10.1038/nphys4091} {\bibfield  {journal} {\bibinfo
  {journal} {Nature Physics}\ }\textbf {\bibinfo {volume} {13}},\ \bibinfo
  {pages} {677} (\bibinfo {year} {2017})}\BibitemShut {NoStop}%
\bibitem [{\citenamefont {Khmel'nitskii}\ and\ \citenamefont
  {Larkin}(1986)}]{khmel_PS1986}%
  \BibitemOpen
  \bibfield  {author} {\bibinfo {author} {\bibfnamefont {D.~E.}\ \bibnamefont
  {Khmel'nitskii}}\ and\ \bibinfo {author} {\bibfnamefont {A.~I.}\ \bibnamefont
  {Larkin}},\ }\bibfield  {title} {\bibinfo {title} {Nonlinear conductance in
  the mesoscopic regime},\ }\href
  {https://doi.org/10.1088/0031-8949/1986/T14/001} {\bibfield  {journal}
  {\bibinfo  {journal} {Physica Scripta}\ }\textbf {\bibinfo {volume} {1986}},\
  \bibinfo {pages} {T14} (\bibinfo {year} {1986})}\BibitemShut {NoStop}%
\bibitem [{\citenamefont {Christen}\ and\ \citenamefont
  {Büttiker}(1996)}]{christen_EPL1996}%
  \BibitemOpen
  \bibfield  {author} {\bibinfo {author} {\bibfnamefont {T.}~\bibnamefont
  {Christen}}\ and\ \bibinfo {author} {\bibfnamefont {M.}~\bibnamefont
  {Büttiker}},\ }\bibfield  {title} {\bibinfo {title} {Gauge-invariant
  nonlinear electric transport in mesoscopic conductors},\ }\href
  {https://doi.org/10.1209/epl/i1996-00145-8} {\bibfield  {journal} {\bibinfo
  {journal} {Europhysics Letters ({EPL})}\ }\textbf {\bibinfo {volume} {35}},\
  \bibinfo {pages} {523} (\bibinfo {year} {1996})}\BibitemShut {NoStop}%
\bibitem [{\citenamefont {Song}\ \emph {et~al.}(1998)\citenamefont {Song},
  \citenamefont {Lorke}, \citenamefont {Kriele}, \citenamefont {Kotthaus},
  \citenamefont {Wegscheider},\ and\ \citenamefont {Bichler}}]{song_PRL1998}%
  \BibitemOpen
  \bibfield  {author} {\bibinfo {author} {\bibfnamefont {A.~M.}\ \bibnamefont
  {Song}}, \bibinfo {author} {\bibfnamefont {A.}~\bibnamefont {Lorke}},
  \bibinfo {author} {\bibfnamefont {A.}~\bibnamefont {Kriele}}, \bibinfo
  {author} {\bibfnamefont {J.~P.}\ \bibnamefont {Kotthaus}}, \bibinfo {author}
  {\bibfnamefont {W.}~\bibnamefont {Wegscheider}},\ and\ \bibinfo {author}
  {\bibfnamefont {M.}~\bibnamefont {Bichler}},\ }\bibfield  {title} {\bibinfo
  {title} {Nonlinear electron transport in an asymmetric microjunction: A
  ballistic rectifier},\ }\href {https://doi.org/10.1103/PhysRevLett.80.3831}
  {\bibfield  {journal} {\bibinfo  {journal} {Phys. Rev. Lett.}\ }\textbf
  {\bibinfo {volume} {80}},\ \bibinfo {pages} {3831} (\bibinfo {year}
  {1998})}\BibitemShut {NoStop}%
\bibitem [{\citenamefont {B{\"u}ttiker}\ and\ \citenamefont
  {Christen}(1998)}]{buttiker_book1998}%
  \BibitemOpen
  \bibfield  {author} {\bibinfo {author} {\bibfnamefont {M.}~\bibnamefont
  {B{\"u}ttiker}}\ and\ \bibinfo {author} {\bibfnamefont {T.}~\bibnamefont
  {Christen}},\ }\href {https://doi.org/10.1007/978-1-4615-5807-1_7} {\emph
  {\bibinfo {title} {Theory of Transport Properties of Semiconductor
  Nanostructures}}},\ edited by\ \bibinfo {editor} {\bibfnamefont
  {E.}~\bibnamefont {Sch{\"o}ll}}\ (\bibinfo  {publisher} {Springer US},\
  \bibinfo {address} {Boston, MA},\ \bibinfo {year} {1998})\ pp.\ \bibinfo
  {pages} {215--248}\BibitemShut {NoStop}%
\bibitem [{\citenamefont {You}\ \emph {et~al.}(2000)\citenamefont {You},
  \citenamefont {Lam},\ and\ \citenamefont {Zheng}}]{you_PRB2000}%
  \BibitemOpen
  \bibfield  {author} {\bibinfo {author} {\bibfnamefont {J.~Q.}\ \bibnamefont
  {You}}, \bibinfo {author} {\bibfnamefont {C.-H.}\ \bibnamefont {Lam}},\ and\
  \bibinfo {author} {\bibfnamefont {H.~Z.}\ \bibnamefont {Zheng}},\ }\bibfield
  {title} {\bibinfo {title} {Landauer-b\"uttiker formula for time-dependent
  transport through resonant-tunneling structures: A nonequilibrium green's
  function approach},\ }\href {https://doi.org/10.1103/PhysRevB.62.1978}
  {\bibfield  {journal} {\bibinfo  {journal} {Phys. Rev. B}\ }\textbf {\bibinfo
  {volume} {62}},\ \bibinfo {pages} {1978} (\bibinfo {year}
  {2000})}\BibitemShut {NoStop}%
\bibitem [{\citenamefont {Shorubalko}\ \emph {et~al.}(2001)\citenamefont
  {Shorubalko}, \citenamefont {Xu}, \citenamefont {Maximov}, \citenamefont
  {Omling}, \citenamefont {Samuelson},\ and\ \citenamefont
  {Seifert}}]{shorubalko_APL2001}%
  \BibitemOpen
  \bibfield  {author} {\bibinfo {author} {\bibfnamefont {I.}~\bibnamefont
  {Shorubalko}}, \bibinfo {author} {\bibfnamefont {H.~Q.}\ \bibnamefont {Xu}},
  \bibinfo {author} {\bibfnamefont {I.}~\bibnamefont {Maximov}}, \bibinfo
  {author} {\bibfnamefont {P.}~\bibnamefont {Omling}}, \bibinfo {author}
  {\bibfnamefont {L.}~\bibnamefont {Samuelson}},\ and\ \bibinfo {author}
  {\bibfnamefont {W.}~\bibnamefont {Seifert}},\ }\bibfield  {title} {\bibinfo
  {title} {Nonlinear operation of \text{GaInAs}/\text{InP}-based three-terminal
  ballistic junctions},\ }\href {https://doi.org/10.1063/1.1396626} {\bibfield
  {journal} {\bibinfo  {journal} {Applied Physics Letters}\ }\textbf {\bibinfo
  {volume} {79}},\ \bibinfo {pages} {1384} (\bibinfo {year}
  {2001})}\BibitemShut {NoStop}%
\bibitem [{\citenamefont {Reitzenstein}\ \emph {et~al.}(2002)\citenamefont
  {Reitzenstein}, \citenamefont {Worschech}, \citenamefont {Hartmann},
  \citenamefont {Kamp},\ and\ \citenamefont {Forchel}}]{reitzenstein_PRL2002}%
  \BibitemOpen
  \bibfield  {author} {\bibinfo {author} {\bibfnamefont {S.}~\bibnamefont
  {Reitzenstein}}, \bibinfo {author} {\bibfnamefont {L.}~\bibnamefont
  {Worschech}}, \bibinfo {author} {\bibfnamefont {P.}~\bibnamefont {Hartmann}},
  \bibinfo {author} {\bibfnamefont {M.}~\bibnamefont {Kamp}},\ and\ \bibinfo
  {author} {\bibfnamefont {A.}~\bibnamefont {Forchel}},\ }\bibfield  {title}
  {\bibinfo {title} {Capacitive-coupling-enhanced switching gain in an electron
  y-branch switch},\ }\href {https://doi.org/10.1103/PhysRevLett.89.226804}
  {\bibfield  {journal} {\bibinfo  {journal} {Phys. Rev. Lett.}\ }\textbf
  {\bibinfo {volume} {89}},\ \bibinfo {pages} {226804} (\bibinfo {year}
  {2002})}\BibitemShut {NoStop}%
\bibitem [{\citenamefont {Polianski}\ and\ \citenamefont
  {B\"uttiker}(2007)}]{polianski_PRB2007}%
  \BibitemOpen
  \bibfield  {author} {\bibinfo {author} {\bibfnamefont {M.~L.}\ \bibnamefont
  {Polianski}}\ and\ \bibinfo {author} {\bibfnamefont {M.}~\bibnamefont
  {B\"uttiker}},\ }\bibfield  {title} {\bibinfo {title} {Rectification and
  nonlinear transport in chaotic dots and rings},\ }\href
  {https://doi.org/10.1103/PhysRevB.76.205308} {\bibfield  {journal} {\bibinfo
  {journal} {Phys. Rev. B}\ }\textbf {\bibinfo {volume} {76}},\ \bibinfo
  {pages} {205308} (\bibinfo {year} {2007})}\BibitemShut {NoStop}%
\bibitem [{\citenamefont {Safi}\ and\ \citenamefont
  {Joyez}(2011)}]{safi_PRB2011}%
  \BibitemOpen
  \bibfield  {author} {\bibinfo {author} {\bibfnamefont {I.}~\bibnamefont
  {Safi}}\ and\ \bibinfo {author} {\bibfnamefont {P.}~\bibnamefont {Joyez}},\
  }\bibfield  {title} {\bibinfo {title} {Time-dependent theory of nonlinear
  response and current fluctuations},\ }\href
  {https://doi.org/10.1103/PhysRevB.84.205129} {\bibfield  {journal} {\bibinfo
  {journal} {Phys. Rev. B}\ }\textbf {\bibinfo {volume} {84}},\ \bibinfo
  {pages} {205129} (\bibinfo {year} {2011})}\BibitemShut {NoStop}%
\bibitem [{\citenamefont {Chang}\ \emph {et~al.}(2015)\citenamefont {Chang},
  \citenamefont {Zhao}, \citenamefont {Kim}, \citenamefont {Wei}, \citenamefont
  {Jain}, \citenamefont {Liu}, \citenamefont {Chan},\ and\ \citenamefont
  {Moodera}}]{chang_PRL2015}%
  \BibitemOpen
  \bibfield  {author} {\bibinfo {author} {\bibfnamefont {C.-Z.}\ \bibnamefont
  {Chang}}, \bibinfo {author} {\bibfnamefont {W.}~\bibnamefont {Zhao}},
  \bibinfo {author} {\bibfnamefont {D.~Y.}\ \bibnamefont {Kim}}, \bibinfo
  {author} {\bibfnamefont {P.}~\bibnamefont {Wei}}, \bibinfo {author}
  {\bibfnamefont {J.~K.}\ \bibnamefont {Jain}}, \bibinfo {author}
  {\bibfnamefont {C.}~\bibnamefont {Liu}}, \bibinfo {author} {\bibfnamefont
  {M.~H.~W.}\ \bibnamefont {Chan}},\ and\ \bibinfo {author} {\bibfnamefont
  {J.~S.}\ \bibnamefont {Moodera}},\ }\bibfield  {title} {\bibinfo {title}
  {Zero-field dissipationless chiral edge transport and the nature of
  dissipation in the quantum anomalous hall state},\ }\href
  {https://doi.org/10.1103/PhysRevLett.115.057206} {\bibfield  {journal}
  {\bibinfo  {journal} {Phys. Rev. Lett.}\ }\textbf {\bibinfo {volume} {115}},\
  \bibinfo {pages} {057206} (\bibinfo {year} {2015})}\BibitemShut {NoStop}%
\bibitem [{\citenamefont {Korniyenko}\ \emph {et~al.}(2016)\citenamefont
  {Korniyenko}, \citenamefont {Shevtsov},\ and\ \citenamefont
  {L\"ofwander}}]{korniyenko_PRB2016}%
  \BibitemOpen
  \bibfield  {author} {\bibinfo {author} {\bibfnamefont {Y.}~\bibnamefont
  {Korniyenko}}, \bibinfo {author} {\bibfnamefont {O.}~\bibnamefont
  {Shevtsov}},\ and\ \bibinfo {author} {\bibfnamefont {T.}~\bibnamefont
  {L\"ofwander}},\ }\bibfield  {title} {\bibinfo {title} {Resonant
  second-harmonic generation in a ballistic graphene transistor with an
  ac-driven gate},\ }\href {https://doi.org/10.1103/PhysRevB.93.035435}
  {\bibfield  {journal} {\bibinfo  {journal} {Phys. Rev. B}\ }\textbf {\bibinfo
  {volume} {93}},\ \bibinfo {pages} {035435} (\bibinfo {year}
  {2016})}\BibitemShut {NoStop}%
\bibitem [{\citenamefont {Rostami}\ and\ \citenamefont
  {Polini}(2018)}]{rostami_PRB2018}%
  \BibitemOpen
  \bibfield  {author} {\bibinfo {author} {\bibfnamefont {H.}~\bibnamefont
  {Rostami}}\ and\ \bibinfo {author} {\bibfnamefont {M.}~\bibnamefont
  {Polini}},\ }\bibfield  {title} {\bibinfo {title} {Nonlinear anomalous
  photocurrents in weyl semimetals},\ }\href
  {https://doi.org/10.1103/PhysRevB.97.195151} {\bibfield  {journal} {\bibinfo
  {journal} {Phys. Rev. B}\ }\textbf {\bibinfo {volume} {97}},\ \bibinfo
  {pages} {195151} (\bibinfo {year} {2018})}\BibitemShut {NoStop}%
\bibitem [{\citenamefont {Texier}\ and\ \citenamefont
  {Mitscherling}(2018)}]{texier_PRB2018}%
  \BibitemOpen
  \bibfield  {author} {\bibinfo {author} {\bibfnamefont {C.}~\bibnamefont
  {Texier}}\ and\ \bibinfo {author} {\bibfnamefont {J.}~\bibnamefont
  {Mitscherling}},\ }\bibfield  {title} {\bibinfo {title} {Nonlinear
  conductance in weakly disordered mesoscopic wires: Interaction and magnetic
  field asymmetry},\ }\href {https://doi.org/10.1103/PhysRevB.97.075306}
  {\bibfield  {journal} {\bibinfo  {journal} {Phys. Rev. B}\ }\textbf {\bibinfo
  {volume} {97}},\ \bibinfo {pages} {075306} (\bibinfo {year}
  {2018})}\BibitemShut {NoStop}%
\bibitem [{\citenamefont {Mardanya}\ \emph {et~al.}(2018)\citenamefont
  {Mardanya}, \citenamefont {Bhattacharya}, \citenamefont {Agarwal},\ and\
  \citenamefont {Dutta}}]{mardanya_PRB2018}%
  \BibitemOpen
  \bibfield  {author} {\bibinfo {author} {\bibfnamefont {S.}~\bibnamefont
  {Mardanya}}, \bibinfo {author} {\bibfnamefont {U.}~\bibnamefont
  {Bhattacharya}}, \bibinfo {author} {\bibfnamefont {A.}~\bibnamefont
  {Agarwal}},\ and\ \bibinfo {author} {\bibfnamefont {A.}~\bibnamefont
  {Dutta}},\ }\bibfield  {title} {\bibinfo {title} {Dynamics of edge currents
  in a linearly quenched haldane model},\ }\href
  {https://doi.org/10.1103/PhysRevB.97.115443} {\bibfield  {journal} {\bibinfo
  {journal} {Phys. Rev. B}\ }\textbf {\bibinfo {volume} {97}},\ \bibinfo
  {pages} {115443} (\bibinfo {year} {2018})}\BibitemShut {NoStop}%
\bibitem [{\citenamefont {Ildarabadi}\ and\ \citenamefont
  {Farghadan}(2021)}]{ildarabadi_PRB2021}%
  \BibitemOpen
  \bibfield  {author} {\bibinfo {author} {\bibfnamefont {F.}~\bibnamefont
  {Ildarabadi}}\ and\ \bibinfo {author} {\bibfnamefont {R.}~\bibnamefont
  {Farghadan}},\ }\bibfield  {title} {\bibinfo {title} {Spin-thermoelectric
  transport in nonuniform strained zigzag graphene nanoribbons},\ }\href
  {https://doi.org/10.1103/PhysRevB.103.115424} {\bibfield  {journal} {\bibinfo
   {journal} {Phys. Rev. B}\ }\textbf {\bibinfo {volume} {103}},\ \bibinfo
  {pages} {115424} (\bibinfo {year} {2021})}\BibitemShut {NoStop}%
\bibitem [{\citenamefont {Beenakker}\ and\ \citenamefont {{van
  Houten}}(1991)}]{bennakker_SSP1991}%
  \BibitemOpen
  \bibfield  {author} {\bibinfo {author} {\bibfnamefont {C.}~\bibnamefont
  {Beenakker}}\ and\ \bibinfo {author} {\bibfnamefont {H.}~\bibnamefont {{van
  Houten}}},\ }\bibfield  {title} {\bibinfo {title} {Quantum transport in
  semiconductor nanostructures},\ }\bibfield  {booktitle} {\emph {\bibinfo
  {booktitle} {Semiconductor Heterostructures and Nanostructures}},\ }\href
  {https://doi.org/https://doi.org/10.1016/S0081-1947(08)60091-0} {\ \bibinfo
  {series} {Solid State Physics},\ \textbf {\bibinfo {volume} {44}},\ \bibinfo
  {pages} {1} (\bibinfo {year} {1991})}\BibitemShut {NoStop}%
\bibitem [{\citenamefont {Altshuler}\ and\ \citenamefont
  {Khmel'nitskii}(1985)}]{altshuler_JETP1985}%
  \BibitemOpen
  \bibfield  {author} {\bibinfo {author} {\bibfnamefont {B.~L.}\ \bibnamefont
  {Altshuler}}\ and\ \bibinfo {author} {\bibfnamefont {D.~E.}\ \bibnamefont
  {Khmel'nitskii}},\ }\bibfield  {title} {\bibinfo {title} {Fluctuation
  properties of small conductors},\ }\href@noop {} {\bibfield  {journal}
  {\bibinfo  {journal} {JETP Letters}\ }\textbf {\bibinfo {volume} {42}},\
  \bibinfo {pages} {359} (\bibinfo {year} {1985})}\BibitemShut {NoStop}%
\bibitem [{\citenamefont {Webb}\ \emph {et~al.}(1988)\citenamefont {Webb},
  \citenamefont {Washburn},\ and\ \citenamefont {Umbach}}]{webb_PRB1988}%
  \BibitemOpen
  \bibfield  {author} {\bibinfo {author} {\bibfnamefont {R.~A.}\ \bibnamefont
  {Webb}}, \bibinfo {author} {\bibfnamefont {S.}~\bibnamefont {Washburn}},\
  and\ \bibinfo {author} {\bibfnamefont {C.~P.}\ \bibnamefont {Umbach}},\
  }\bibfield  {title} {\bibinfo {title} {Experimental study of nonlinear
  conductance in small metallic samples},\ }\href
  {https://doi.org/10.1103/PhysRevB.37.8455} {\bibfield  {journal} {\bibinfo
  {journal} {Phys. Rev. B}\ }\textbf {\bibinfo {volume} {37}},\ \bibinfo
  {pages} {8455} (\bibinfo {year} {1988})}\BibitemShut {NoStop}%
\bibitem [{\citenamefont {Buttiker}(1993)}]{buttiker_JPCM1993}%
  \BibitemOpen
  \bibfield  {author} {\bibinfo {author} {\bibfnamefont {M.}~\bibnamefont
  {Buttiker}},\ }\bibfield  {title} {\bibinfo {title} {Capacitance, admittance,
  and rectification properties of small conductors},\ }\href
  {https://doi.org/10.1088/0953-8984/5/50/017} {\bibfield  {journal} {\bibinfo
  {journal} {Journal of Physics: Condensed Matter}\ }\textbf {\bibinfo {volume}
  {5}},\ \bibinfo {pages} {9361} (\bibinfo {year} {1993})}\BibitemShut
  {NoStop}%
\bibitem [{\citenamefont {Thakur}\ \emph {et~al.}(2004)\citenamefont {Thakur},
  \citenamefont {Green},\ and\ \citenamefont {Das}}]{thakur_IJMPB2004}%
  \BibitemOpen
  \bibfield  {author} {\bibinfo {author} {\bibfnamefont {J.~S.}\ \bibnamefont
  {Thakur}}, \bibinfo {author} {\bibfnamefont {F.}~\bibnamefont {Green}},\ and\
  \bibinfo {author} {\bibfnamefont {M.~P.}\ \bibnamefont {Das}},\ }\bibfield
  {title} {\bibinfo {title} {Sum-rule constraints for open mesoscopic
  conductors},\ }\href {https://doi.org/10.1142/S0217979204024938} {\bibfield
  {journal} {\bibinfo  {journal} {International Journal of Modern Physics B}\
  }\textbf {\bibinfo {volume} {18}},\ \bibinfo {pages} {1479} (\bibinfo {year}
  {2004})}\BibitemShut {NoStop}%
\bibitem [{\citenamefont {S\'anchez}\ and\ \citenamefont
  {B\"uttiker}(2004)}]{david_PRL2004}%
  \BibitemOpen
  \bibfield  {author} {\bibinfo {author} {\bibfnamefont {D.}~\bibnamefont
  {S\'anchez}}\ and\ \bibinfo {author} {\bibfnamefont {M.}~\bibnamefont
  {B\"uttiker}},\ }\bibfield  {title} {\bibinfo {title} {Magnetic-field
  asymmetry of nonlinear mesoscopic transport},\ }\href
  {https://doi.org/10.1103/PhysRevLett.93.106802} {\bibfield  {journal}
  {\bibinfo  {journal} {Phys. Rev. Lett.}\ }\textbf {\bibinfo {volume} {93}},\
  \bibinfo {pages} {106802} (\bibinfo {year} {2004})}\BibitemShut {NoStop}%
\bibitem [{\citenamefont {L\"ofgren}\ \emph {et~al.}(2004)\citenamefont
  {L\"ofgren}, \citenamefont {Marlow}, \citenamefont {Shorubalko},
  \citenamefont {Taylor}, \citenamefont {Omling}, \citenamefont {Samuelson},\
  and\ \citenamefont {Linke}}]{lofgren_PRL2004}%
  \BibitemOpen
  \bibfield  {author} {\bibinfo {author} {\bibfnamefont {A.}~\bibnamefont
  {L\"ofgren}}, \bibinfo {author} {\bibfnamefont {C.~A.}\ \bibnamefont
  {Marlow}}, \bibinfo {author} {\bibfnamefont {I.}~\bibnamefont {Shorubalko}},
  \bibinfo {author} {\bibfnamefont {R.~P.}\ \bibnamefont {Taylor}}, \bibinfo
  {author} {\bibfnamefont {P.}~\bibnamefont {Omling}}, \bibinfo {author}
  {\bibfnamefont {L.}~\bibnamefont {Samuelson}},\ and\ \bibinfo {author}
  {\bibfnamefont {H.}~\bibnamefont {Linke}},\ }\bibfield  {title} {\bibinfo
  {title} {Symmetry of two-terminal nonlinear electric conduction},\ }\href
  {https://doi.org/10.1103/PhysRevLett.92.046803} {\bibfield  {journal}
  {\bibinfo  {journal} {Phys. Rev. Lett.}\ }\textbf {\bibinfo {volume} {92}},\
  \bibinfo {pages} {046803} (\bibinfo {year} {2004})}\BibitemShut {NoStop}%
\bibitem [{\citenamefont {Datta}(2017)}]{datta_book2}%
  \BibitemOpen
  \bibfield  {author} {\bibinfo {author} {\bibfnamefont {S.}~\bibnamefont
  {Datta}},\ }\href {https://books.google.co.in/books?id=eNOkAQAACAAJ} {\emph
  {\bibinfo {title} {Lessons from Nanoelectronics: A New Perspective on
  Transport}}},\ \bibinfo {series} {Lessons from nanoscience}\ No.\ \bibinfo
  {number} {pt. 1}\ (\bibinfo  {publisher} {World Scientific},\ \bibinfo {year}
  {2017})\BibitemShut {NoStop}%
\bibitem [{\citenamefont {Shi}\ and\ \citenamefont {Song}(2019)}]{shi_PRB2019}%
  \BibitemOpen
  \bibfield  {author} {\bibinfo {author} {\bibfnamefont {L.-k.}\ \bibnamefont
  {Shi}}\ and\ \bibinfo {author} {\bibfnamefont {J.~C.~W.}\ \bibnamefont
  {Song}},\ }\bibfield  {title} {\bibinfo {title} {Symmetry, spin-texture, and
  tunable quantum geometry in a \text{WTe}$_{2}$ monolayer},\ }\href
  {https://doi.org/10.1103/PhysRevB.99.035403} {\bibfield  {journal} {\bibinfo
  {journal} {Phys. Rev. B}\ }\textbf {\bibinfo {volume} {99}},\ \bibinfo
  {pages} {035403} (\bibinfo {year} {2019})}\BibitemShut {NoStop}%
\bibitem [{\citenamefont {Sun}\ \emph {et~al.}(2020)\citenamefont {Sun},
  \citenamefont {Song}, \citenamefont {Weng},\ and\ \citenamefont
  {Dai}}]{sun_PRB2020}%
  \BibitemOpen
  \bibfield  {author} {\bibinfo {author} {\bibfnamefont {S.}~\bibnamefont
  {Sun}}, \bibinfo {author} {\bibfnamefont {Z.}~\bibnamefont {Song}}, \bibinfo
  {author} {\bibfnamefont {H.}~\bibnamefont {Weng}},\ and\ \bibinfo {author}
  {\bibfnamefont {X.}~\bibnamefont {Dai}},\ }\bibfield  {title} {\bibinfo
  {title} {Topological metals induced by the zeeman effect},\ }\href
  {https://doi.org/10.1103/PhysRevB.101.125118} {\bibfield  {journal} {\bibinfo
   {journal} {Phys. Rev. B}\ }\textbf {\bibinfo {volume} {101}},\ \bibinfo
  {pages} {125118} (\bibinfo {year} {2020})}\BibitemShut {NoStop}%
\bibitem [{\citenamefont {Morimoto}\ and\ \citenamefont
  {Nagaosa}(2018)}]{morimoto_SR2018}%
  \BibitemOpen
  \bibfield  {author} {\bibinfo {author} {\bibfnamefont {T.}~\bibnamefont
  {Morimoto}}\ and\ \bibinfo {author} {\bibfnamefont {N.}~\bibnamefont
  {Nagaosa}},\ }\bibfield  {title} {\bibinfo {title} {Nonreciprocal current
  from electron interactions in noncentrosymmetric crystals: roles of time
  reversal symmetry and dissipation},\ }\href
  {https://doi.org/10.1038/s41598-018-20539-2} {\bibfield  {journal} {\bibinfo
  {journal} {Scientific Reports}\ }\textbf {\bibinfo {volume} {8}},\ \bibinfo
  {pages} {2973} (\bibinfo {year} {2018})}\BibitemShut {NoStop}%
\bibitem [{\citenamefont {Arora}\ \emph {et~al.}(2020)\citenamefont {Arora},
  \citenamefont {Shi},\ and\ \citenamefont {Song}}]{arpit_PRB2020}%
  \BibitemOpen
  \bibfield  {author} {\bibinfo {author} {\bibfnamefont {A.}~\bibnamefont
  {Arora}}, \bibinfo {author} {\bibfnamefont {L.-k.}\ \bibnamefont {Shi}},\
  and\ \bibinfo {author} {\bibfnamefont {J.~C.~W.}\ \bibnamefont {Song}},\
  }\bibfield  {title} {\bibinfo {title} {Cooperative orbital moments and edge
  magnetoresistance in monolayer \text{WTe}$_{2}$},\ }\href
  {https://doi.org/10.1103/PhysRevB.102.161402} {\bibfield  {journal} {\bibinfo
   {journal} {Phys. Rev. B}\ }\textbf {\bibinfo {volume} {102}},\ \bibinfo
  {pages} {161402} (\bibinfo {year} {2020})}\BibitemShut {NoStop}%
\bibitem [{\citenamefont {Fu}(2009)}]{Fu_PRL2009}%
  \BibitemOpen
  \bibfield  {author} {\bibinfo {author} {\bibfnamefont {L.}~\bibnamefont
  {Fu}},\ }\bibfield  {title} {\bibinfo {title} {Hexagonal warping effects in
  the surface states of the topological insulator
  \text{Bi}$_{2}$\text{Te}$_{3}$},\ }\href
  {https://doi.org/10.1103/PhysRevLett.103.266801} {\bibfield  {journal}
  {\bibinfo  {journal} {Phys. Rev. Lett.}\ }\textbf {\bibinfo {volume} {103}},\
  \bibinfo {pages} {266801} (\bibinfo {year} {2009})}\BibitemShut {NoStop}%
\bibitem [{\citenamefont {Durnev}\ and\ \citenamefont
  {Tarasenko}(2016)}]{durnev_PRB2016}%
  \BibitemOpen
  \bibfield  {author} {\bibinfo {author} {\bibfnamefont {M.~V.}\ \bibnamefont
  {Durnev}}\ and\ \bibinfo {author} {\bibfnamefont {S.~A.}\ \bibnamefont
  {Tarasenko}},\ }\bibfield  {title} {\bibinfo {title} {Magnetic field effects
  on edge and bulk states in topological insulators based on
  \text{HgTe}/\text{CdHgTe} quantum wells with strong natural interface
  inversion asymmetry},\ }\href {https://doi.org/10.1103/PhysRevB.93.075434}
  {\bibfield  {journal} {\bibinfo  {journal} {Phys. Rev. B}\ }\textbf {\bibinfo
  {volume} {93}},\ \bibinfo {pages} {075434} (\bibinfo {year}
  {2016})}\BibitemShut {NoStop}%
\end{thebibliography}
\end{document}